\documentclass{article}%
\usepackage{amssymb}
\usepackage{amsfonts}
\usepackage{amsmath}
\usepackage{geometry}
\usepackage{scalefnt}
\usepackage{theorem}
\usepackage{graphicx}%
{\theorembodyfont{\upshape}

}

{\theorembodyfont{\upshape}

}

{\theorembodyfont{\upshape}

}

\begin{document}

\title{Estimation of pure qubits on circles}
\author{Samuel L. Braunstein\thanks{ Department of Computer Science, University of
York, \ Heslington, York YO10 5DD, United Kingdom; schmuel@cs.york.ac.uk}
\and Sibasish Ghosh\thanks{Department of Computer Science, University of York;
sibasish@cs.york.ac.uk}
\and Simone Severini\thanks{Department of Mathematics and Department of Computer
Science, University of York; ss54@york.ac.uk}}
\maketitle

\begin{abstract}
Gisin and Popescu [PRL, 83, 432 (1999)] have shown that more information about
their direction can be obtained from a pair of anti-parallel spins compared to
a pair of parallel spins, where the first member of the pair (which we call
the pointer member) can point equally along any direction in the Bloch sphere.
They argued that this was due to the difference in dimensionality spanned by
these two alphabets of states. Here we consider similar alphabets, but with
the first spin restricted to a fixed small circle of the Bloch sphere. In this
case, the dimensionality spanned by the anti-parallel versus parallel alphabet
is now equal. However, the anti-parallel alphabet is found to still contain
more information in general. We generalize this to having $N$ parallel spins
and $M$ anti-parallel spins. When the pointer member is restricted to a small
circle these alphabets again span spaces of equal dimension, yet in general,
more directional information can be found for sets with smaller $|N-M|$ for
any fixed total number of spins. We find that the optimal POVMs for extracting
directional information in these cases can always be expressed in terms of the
Fourier basis. Our results show that dimensionality alone cannot explain the
greater information content in anti-parallel combinations of spins compared to
parallel combinations. In addition, we describe an LOCC protocol which extract
optimal directional information when the pointer member is restricted to a
small circle and a pair of parallel spins are supplied.

\end{abstract}
\tableofcontents

\section{Introduction}

In the quantum world there are many phenomena whose explanation is beyond the
intuitions suggested by the classical world. For example impossibility of
\emph{cloning} \cite{wz} or \emph{deleting} \cite{sp} an arbitrary quantum
state, existence of \emph{non-orthogonal} quantum states \cite{p},
impossibility of \emph{spin-flipping }\cite{gisinpopescu99}, \emph{etc.} Let
us focus on this last concept. The spin degree of freedom of a spin $1/2$
system is described by a vector (or a mixture of projections on vectors) of a
$2$-dimensional Hilbert space $\mathcal{H}$. The quantum mechanical operation
which, when applied to a qubit $|\psi\rangle\in\mathcal{H}$, produces the
qubit $|\psi^{\perp}\rangle\in\mathcal{H}$, orthogonal to $|\psi\rangle$, is
called \emph{spin-flipping}. This operation exists if and only if the Bloch
vector of $|\psi\rangle$ lies on \emph{a given} great circle \cite{g}. The
classical analogue of spin-flipping is the operation which, when applied to a
vector (for example, in Eucledian space), produces its negative. This
operation always exists.

It is provable that we can extract more information about the direction of the
Bloch vector of $|\psi\rangle$ (for short, the \emph{direction} of
$|\psi\rangle$) if, instead of just $|\psi\rangle$, we are supplied with
$|\psi\rangle\otimes|\psi\rangle$ or $|\psi\rangle\otimes|\psi^{\perp}\rangle
$, in words, a pair of \emph{parallel} or \emph{anti-parallel} qubits. In a
classical scenario, there is no difference between the parallel and the
anti-parallel case, since the classical analogue of spin-flipping always
exists. Gisin and Popescu \cite{gisinpopescu99} have shown that, when
concerning the direction of $|\psi\rangle$, anti-parallel qubits provide more
information than parallel qubits. Notice that if spin-flipping were possible
for each vector of $\mathcal{H}$, there would be no difference between these
two cases. Gisin and Popescu pointed out that parallel qubits span a
$3$-dimensional subspace of $\mathcal{H}\otimes\mathcal{H}$, while
anti-parallel qubits span $\mathcal{H}\otimes\mathcal{H}$ entirely.
Intuitively, vectors in an enlarged space are better distinguished than in the
original space; the better we can distinguish the parallel (or the
anti-parallel) qubits, the more information we can extract about the direction
of $|\psi\rangle$. So, according to Gisin and Popescu, anti-parallel qubits
contain more information about the direction of $|\psi\rangle$ compared to
parallel qubits, since the former span a space of higher dimension.

The following question arises naturally: does this difference in extracting
information about the direction of a qubit occur only for parallel and
anti-parallel qubits? Can this be generalized to other cases? Let us
illustrate the situation for general operations rather than just
spin-flipping. Consider an operation (not necessarily quantum mechanical)
$\mathcal{A}$ on the pure state $|\psi\rangle$ of a spin $1/2$ system, such
that $|\langle\psi|\mathcal{A}|\psi\rangle|$ is independent of $|\psi\rangle$.
It can be shown that no non-trivial quantum mechanical operation satisfies
this last requirement. In particular, spin-flipping $\mathcal{A}$ should be of
the form $\mathcal{A}|\psi\rangle=|\psi^{\perp}\rangle$, where $\langle
\psi^{\perp}|\psi\rangle=0$. One can raise now another question: which of the
following two sets $\{|\psi\rangle\otimes\mathcal{A}|\psi\rangle:|\psi\rangle$
is any normalized qubit$\}$ and $\{|\psi\rangle\otimes\mathcal{B}|\psi
\rangle:|\psi\rangle$ is any normalized qubit$\}$, where $\mathcal{A}$ and
$\mathcal{B}$ are two operations such that both $|\langle\psi|\mathcal{A}%
|\psi\rangle|$ and $|\langle\psi|\mathcal{B}|\psi\rangle|$ are independent of
$|\psi\rangle$, contains more information about the direction of the qubit?
The problem can be posed in the following more general form. Let
$f_{i}:\mathcal{H}\longrightarrow\mathcal{H}^{\otimes n}$, for $i=1,2,...$, be
one-to-one maps which take $|\psi\rangle$ into some state of $\mathcal{H}%
^{\otimes n}$. Which $f_{i}$ gives the largest amount of information about the
direction of $|\psi\rangle$, when $|\psi\rangle$ is sampled according to some
\emph{a priori} probability distribution on $\mathcal{H}$? The maps $f_{i}$'s
are said to provide an \emph{encoding} of vectors in $\mathcal{H}$. Note that
spin-flipping provides an encoding.

Bagan \emph{et al.} \cite{bagan02} have discussed this problem considering
only those maps for which $f_{i}(|\psi\rangle)$ is an eigenstate of a total
spin observable along some direction (specified by the direction of
$|\psi\rangle$), when $|\psi\rangle$ is sampled from the uniform distribution
on $\mathcal{H}$. According to their analysis, spin-flipping does not play any
special role; all that matters (in order to extract information) is the
dimension of the subspace spanned by the states $f_{i}(|\psi\rangle)$.

\bigskip

In the present paper, we consider this \emph{dimensional argument} in the
context of estimating the direction of a pure qubit when the state
$f_{i}(|\psi\rangle)$ is of the form $|\psi\rangle^{\otimes n}\otimes
|\psi^{\prime}\rangle^{\otimes m}$. Here the qubits $|\psi\rangle$ and
$|\psi^{\prime}\rangle$ are in one-to-one correspondence. Moreover we assume
that the Bloch vectors of $|\psi\rangle$ and $|\psi^{\prime}\rangle$ lie on
two fixed different circles (possibly small). What is the motivation behind
the choice of this encoding? First of all, small circles are a plausible first
step generalization of great circles even though spin-flipping does not exist
for states from small circles. However, $|\psi\rangle\otimes|\psi\rangle$ and
$|\psi\rangle\otimes|\psi^{\perp}\rangle$, where the Bloch vector of
$|\psi\rangle$ lies on a given small circle, span $3$-dimensional subspaces.
Thus, in our framework, the dimensional arguments of Gisin-Popescu and Bagan
\emph{et al.} do not give any clue regarding best extraction directional
information of the qubit. We find that, even in this case, anti-parallel
qubits contain more information compared to parallel ones. More generally we
see that the information contained in $N$ qubits about the direction of
$|\psi\rangle$, when $|\psi\rangle$ is encoded in the state $|\psi
\rangle^{\otimes n}\otimes|\psi^{\perp}\rangle^{\otimes(N-n)}$ and taken with
equal probability from a given small circle, decreases with the increment of
the difference $|N-2n|$.

We next consider the problem of estimating the direction of $|\psi\rangle$
with an encoding of the form $|\psi\rangle\otimes|\psi^{\prime}\rangle$, where
the Bloch vectors of $|\psi\rangle$ and $|\psi^{\prime}\rangle$ lie
respectively on two parallel circles $S$ and $S^{\prime}$, and $|\psi^{\prime
}\rangle$ is in one-to-one correspondence with $|\psi\rangle$. Note that
encoding parallel and anti-parallel qubits are special cases of the this type
of encoding. Given two circles $S$ and $S^{\prime}$, we have an expression for
$F(S,S^{\prime})$, the maximum amount of information about the direction of
$|\psi\rangle$. Given a circle $S$, one can calculate the maximum and minimum
value of $F(S,S^{\prime})$ over all possible choices of $S^{\prime}$. Let us
denote by $F^{\max}(S)$ and $F^{\min}(S)$ the maximum and the minimum value of
$F(S,S^{\prime})$, respectively. We show that $F^{\max}(S)\geq F(S,S^{\perp
})\geq F(S,S)\geq F^{\min}(S)$ for all $S$, where $S^{\perp}$ is the circle
diametrically opposite to $S$. Also in this case the dimensional argument does
not work. Another scenario in which the dimensional argument fails is the case
of estimating the direction of a qubit sampled from a uniform distribution on
set of two diametrically opposite circles. Here parallel and anti-parallel
qubits provide same information about the direction of the qubit, even though
they span spaces of different dimension.

Estimating a qubit from a circle is essentially estimating the phase of the
qubit (see Section 4 below and, \emph{e.g.}, \cite{d}). Bearing this in mind,
one can argue that the measurement basis used in the optimal estimation
strategy should be the Fourier basis. This is also reflected by our results.
In fact, we find that the measurement basis for the optimal strategy in the
case of a qubit $|\psi\rangle$ uniformly distributed on a small circle,
supplied the state $|\psi\rangle^{\otimes n}\otimes|\psi^{\perp}%
\rangle^{\otimes(N-n)}$ (where $N$ is fixed), is the $(N+1)$-dimensional
Fourier basis for every $n\in\{1,2,...,N-1\}$. This is also true if the
supplied state is $|\psi\rangle^{\otimes n}\otimes|\psi^{\prime}\rangle$,
where $|\psi\rangle$ and $|\psi^{\prime}\rangle$ are respectively from two
different parallel circles and in one-to-one correspondence. This is expected
as in the case of phase estimation. Moreover, we see that the measurement
basis for the optimal strategy in the case of a qubit $|\psi\rangle$,
uniformly distributed on a set of two diametrically opposite circles, supplied
the state $|\psi\rangle\otimes|\psi\rangle$, is again the Fourier basis (in
three dimensions). It is remarkable that this scenario does not correspond to
phase estimation. Finally, we observe that, if the supplied state is
$|\psi\rangle\otimes|\psi^{\perp}\rangle$ then the measurement basis of the
optimal strategy is not the Fourier basis, but is in some way
\textquotedblleft similar\textquotedblright\ to the Haar basis (see,
\emph{e.g.}, \cite{da} for this notion).

\bigskip

The paper is organized as follows. In Section 2 we formulate the problem of
state estimation discussed in the paper. In Section 3 we sketch the related
previous works. In Section 4 we tackle the problem of estimating the direction
of the Bloch vector of a qubit $|\psi\rangle$ sampled from a uniform
distribution on a given small circle, when are supplied states of the form
$|\psi\rangle^{\otimes n}\otimes|\psi^{\perp}\rangle^{\otimes m}$. In Section
5 we consider the case in which the supplied states are of the form
$|\psi\rangle\otimes|\psi^{\prime}\rangle$, where $|\psi\rangle$ and
$|\psi^{\prime}\rangle$ are respectively from two parallel circles and in
one-to-one correspondence with each other. In Section 6 we consider the case
of a qubit $|\psi\rangle$ sampled from a uniform distribution on two
diametrically opposite circles and the supplied states are either
$|\psi\rangle\otimes|\psi\rangle$ or $|\psi\rangle\otimes|\psi^{\perp}\rangle
$. In Section 7 we describe an LOCC protocol for optimally estimating the
direction of a qubit $\left\vert \psi\left(  \theta,\phi\right)  \right\rangle
$ ($\theta$ is fixed), when a pair of parallel qubits is supplied. Section 8
is devoted to discussion and open problems.

\section{The problem of state estimation}

\subsection{Formulation}

Let us consider a quantum mechanical system with associated Hilbert space
$\mathcal{H}\cong\mathbb{C}^{d}$. Let $A$ be a set of indices (not necessarily
countable) and let $S=\{|\psi_{\alpha}\rangle:\alpha\in A\}\subseteq
\mathcal{H}$ be a set of normalized pure states. This is equivalent to say
that each $|\psi_{\alpha}\rangle\in S$ is of the form $|\psi_{\alpha}%
\rangle=\sum_{i=1}^{d}\psi_{\alpha_{i}}|\psi_{i}\rangle$, where $\{|\psi
_{1}\rangle,|\psi_{2}\rangle,...,|\psi_{d}\rangle\}$ is an orthonormal basis
of $\mathcal{H}$ and $\psi_{\alpha_{1}},\psi_{\alpha_{2}},...,\psi_{\alpha
_{d}}$ are complex numbers such that $\sum_{i=1}^{d}|\psi_{\alpha_{i}}|^{2}%
=1$. Suppose that we want to gather information about an unknown state
$|\psi_{x}\rangle\in S$. Once we have chosen, and fixed, an orthonormal basis
of $\mathcal{H}$, say $\{|\psi_{1}\rangle,|\psi_{2}\rangle,...,|\psi
_{d}\rangle\}$, information about the coefficients $\psi_{x_{1}},\psi_{x_{2}%
},...,\psi_{x_{d}}$ is obtained by performing measurements on the state
$|\psi_{x}\rangle$. The mathematical description of a general measurement on a
quantum state is the \emph{Positive Operator Valued Measurement} (POVM)
formalism. This is described as follows. Let $\Lambda$ be a set of indices
(not necessarily finite). A POVM $\mathcal{M}=\{\widehat{E}_{r}:r\in\Lambda\}$
on $\mathcal{H}$ is a set of positive operators $\widehat{E}_{r}%
:\mathcal{H}\longrightarrow\mathcal{H}$ such that $\sum_{r\in\Lambda}%
\widehat{E}_{r}=\widehat{I}_{\mathcal{H}}$, where $\widehat{I}_{\mathcal{H}}$
is the identity operator in $\mathcal{H}$. The probability that the $r$-th
measurement outcome occurs is given by $\langle\psi_{x}|\widehat{E}_{r}%
|\psi_{x}\rangle$ and the state immediately after the measurement is
$(\langle\psi_{x}|\widehat{E}_{r}^{\dagger}\widehat{E}_{r}|\psi_{x}%
\rangle)^{-\frac{1}{2}}\widehat{E}_{r}|\psi_{x}\rangle$ (where $|\psi
_{x}\rangle$ is the state before the measurement). For all practical purposes
$\Lambda$ is taken to be finite. In such a case, since the Hilbert space
$\mathcal{H}$ is finite dimensional, it follows by a theorem of Davies
\cite{dav} that the elements of $\mathcal{M}$ can be chosen to be of rank one.
In this paper, we consider POVMs with rank one elements only. As a matter of
fact full information about $\psi_{x_{1}},\psi_{x_{2}},...,\psi_{x_{d}}$ is
obtained only by performing measurements on an infinite number of copies of
$|\psi_{x}\rangle$. Since it is physically impossible to be supplied with an
infinite number of copies of a quantum state, we assume that we are supplied
with $n$ copies of $|\psi_{x}\rangle$ only. Now, let $A^{\prime}$ be a set of
indices (not necessarily countable) and let $S_{n}=\{|\Psi_{\alpha^{\prime}%
}\rangle:\alpha^{\prime}\in A^{\prime}\}\subseteq\mathcal{H}^{\otimes n}$ be a
set of normalized pure states in $\mathcal{H}^{\otimes n}$. We assume that
there is a bijective function $f:A\longrightarrow A^{\prime}$. A \emph{state
estimation strategy} $(\mathcal{M},T)$ is composed of:

\begin{itemize}
\item A POVM on $\mathcal{H}^{\otimes n}$, $\mathcal{M}=\{\widehat{E}_{r}%
:r\in\Lambda\}$;

\item A set of density matrices $T=\{\rho_{r}:r\in\Lambda\}\subseteq S$.
\end{itemize}

\noindent If the $r$-th outcome of a measurement performed by applying
$\mathcal{M}$ to a given $|\Psi_{\alpha^{\prime}}\rangle\in S_{n}$ occurs, the
system is then prepared in the state $\rho_{r}$. This is said to be the
\emph{estimated state} from the $r$-th measurement outcome. For any
$|\Psi_{\alpha^{\prime}}\rangle\in S_{n}$, the \emph{average estimated state}
of the system is given by%
\[
\rho^{(\Psi_{\alpha^{\prime}})}:=\sum\limits_{r\in\Lambda}\langle\Psi
_{\alpha^{\prime}}|\widehat{E}_{r}|\Psi_{\alpha^{\prime}}\rangle\rho_{r}.
\]
Thus, the \emph{fidelity} for $(\mathcal{M},T)$ to estimate the state
$|\psi_{f^{-1}(\alpha^{\prime})}\rangle$ is $\langle\psi_{f^{-1}%
(\alpha^{\prime})}|\rho^{(\Psi_{\alpha^{\prime}})}|\psi_{f^{-1}(\alpha
^{\prime})}\rangle$, and the \emph{average fidelity} for $(\mathcal{M},T)$ to
estimate states in $S$ is given by
\begin{equation}
\overline{F}(\mathcal{M},T):=\int\limits_{\alpha^{\prime}\in A^{\prime}%
}\langle\psi_{f^{-1}(\alpha^{\prime})}|\rho^{(\Psi_{\alpha^{\prime}})}%
|\psi_{f^{-1}(\alpha^{\prime})}\rangle d(\alpha^{\prime}), \label{e0}%
\end{equation}
where $d(\alpha^{\prime})$ is a generalized measure over $A^{\prime}$. Then,
using the above expression for $\rho^{(\Psi_{\alpha^{\prime}})}$, we have%
\begin{equation}
\overline{F}(\mathcal{M},T)=\int\limits_{\alpha^{\prime}\in A^{\prime}}%
\sum\limits_{r\in\Lambda}\langle\Psi_{\alpha^{\prime}}|\widehat{E}_{r}%
|\Psi_{\alpha^{\prime}}\rangle\langle\psi_{f^{-1}(\alpha^{\prime})}|\rho
_{r}|\psi_{f^{-1}(\alpha^{\prime})}\rangle d(\alpha^{\prime}). \label{e1}%
\end{equation}
In this last equation, we can replace $\langle\psi_{f^{-1}(\alpha^{\prime}%
)}|\rho_{r}|\psi_{f^{-1}(\alpha^{\prime})}\rangle$ with a $[0,1]$-valued
parameter depending on $\mathcal{M},r$ and $\alpha^{\prime}$. Such a
parameter, denoted by $s(\mathcal{M},r,\alpha^{\prime})$, is called the
\emph{score}. We write $\overline{F}(\mathcal{M},s)$ if we want to stress that
the average fidelity depends also on a specified score $s(\mathcal{M}%
,r,\alpha^{\prime})$. Notice that $T$ depends on the choice of $s$. Let
\[
\overline{F}_{s}^{\max}:=\sup_{\mathcal{M}}\overline{F}(\mathcal{M},s).
\]
Our task is to evaluate $\overline{F}_{s}^{\max}$ and to determine which state
estimation strategies achieve this quantity. We simply write $\overline
{F}^{\max}$ if the score is taken to be $\langle\psi_{f^{-1}(\alpha^{\prime}%
)}|\rho_{r}|\psi_{f^{-1}(\alpha^{\prime})}\rangle$.

\subsection{How to evaluate $\overline{F}_{s}^{\max}$}

Let us denote by $\mathcal{L}(A)$ the linear span of a set of vectors $A$. Let
$\{|\Psi_{i}\rangle:i=1,...,N\}$ be an orthonormal basis of $\mathcal{L}%
(S_{n})$. We can attain higher values of $\overline{F}(\mathcal{M},s)$ if we
restrict each POVM element $\widehat{E}_{r}$ to have support in $\mathcal{L}%
(S_{n})$ instead of $\mathcal{H}^{\otimes n}$ (recall that the \emph{support}
of an operator is the linear span of its range). In fact, since $|\Psi
_{\alpha^{\prime}}\rangle\in\mathcal{L}(S_{n})$, one can get higher value of
$\langle\Psi_{\alpha^{\prime}}|\widehat{E}_{r}|\Psi_{\alpha^{\prime}}\rangle$
if every $\widehat{E}_{r}$ has support in $\mathcal{L}(S_{n})$. However, note
that if the elements of a POVM have support in a subspace of $\mathcal{H}%
^{\otimes n}$ containing $\mathcal{L}(S_{n})$, then the POVM may still give
rise to $\overline{F}_{s}^{\max}$ (an example is given in Section \ref{s}
below). In order to compute $\overline{F}(\mathcal{M},s)$, we take the POVM
$\mathcal{M}=\{\widehat{E}_{r}:r\in\Lambda\}$ such that, for every
$r\in\Lambda$, we have $\widehat{E}_{r}=C_{r}P\left[  \sum_{i=1}^{N}%
\lambda_{ir}|\Psi_{i}\rangle\right]  $, with the following constraints:

\begin{enumerate}
\item[A.] $C_{r}>0$ for every $r\in\Lambda$;

\item[B.] $\sum_{i=1}^{N}|\lambda_{ir}|^{2}=1$, for every $r\in\Lambda$.
\end{enumerate}

Then each $\widehat{E}_{r}$ has support in $\mathcal{L}(S_{n})$. These
operators form a POVM on $\mathcal{L}(S_{n})$ if and only if%
\[%
\begin{tabular}
[c]{lll}%
$\sum\limits_{r\in\Lambda}C_{r}P\left[  \sum\limits_{i=1}^{N}\lambda_{ir}%
|\Psi_{i}\rangle\right]  =I_{\mathcal{L}(S_{n})}=\sum\limits_{i=1}^{N}%
P[|\Psi_{i}\rangle],$ &  & that is, if and only if:
\end{tabular}
\]

\begin{enumerate}
\item[C.] $\sum_{r\in\Lambda}C_{r}\lambda_{ir}\lambda_{jr}^{\ast}=\delta_{ij}$
for every $i,j=1,...,N$.
\end{enumerate}

Now, the given state of $S_{n}$ can be written as $|\Psi_{\alpha^{\prime}%
}\rangle=\sum_{i=1}^{N}\mu_{i}(\alpha^{\prime})|\Psi_{i}\rangle$, where:

\begin{enumerate}
\item[D.] $\sum_{i=1}^{N}|\mu_{i}(\alpha^{\prime})|^{2}=1$.
\end{enumerate}

Note that, although the set $\{(\mu_{1}(\alpha^{\prime}),\mu_{2}%
(\alpha^{\prime}),...,\mu_{N}(\alpha^{\prime})):\alpha^{\prime}\in A^{\prime
}\}$ is known, the individual $N$-tuple $(\mu_{1}(\alpha^{\prime}),\mu
_{2}(\alpha^{\prime}),...,\mu_{N}(\alpha^{\prime}))$ is not, since the
supplied state $|\Psi_{\alpha^{\prime}}\rangle$ is unknown. Thus
\[%
\begin{tabular}
[c]{lll}%
$\langle\Psi_{\alpha^{\prime}}|\widehat{E}_{r}|\Psi_{\alpha^{\prime}}%
\rangle=C_{r}\left\vert \sum\limits_{j=1}^{N}\lambda_{jr}^{\ast}\mu_{i}%
(\alpha^{\prime})\right\vert ^{2},$ &  & $\text{for every }r\in\Lambda.$%
\end{tabular}
\]
It follows that, with the score $\langle\psi_{f^{-1}(\alpha^{\prime})}%
|\rho_{r}|\psi_{f^{-1}(\alpha^{\prime})}\rangle$, the average fidelity is%
\[
\overline{F}(\mathcal{M},T)=\int\limits_{\alpha^{\prime}\in A^{\prime}}\left(
\sum\limits_{r\in\Lambda}C_{r}\left\vert \sum\limits_{j=1}^{N}\lambda
_{jr}^{\ast}\mu_{i}(\alpha^{\prime})\right\vert ^{2}\langle\psi_{f^{-1}%
(\alpha^{\prime})}|\rho_{r}|\psi_{f^{-1}(\alpha^{\prime})}\rangle\right)
d(\alpha^{\prime})
\]
or, equivalently,
\begin{equation}
\overline{F}(\mathcal{M},T)=\sum\limits_{r\in\Lambda}\sum\limits_{j,k=1}%
^{N}C_{r}\lambda_{jr}^{\ast}\lambda_{kr}\left(  \int\limits_{\alpha^{\prime
}\in A^{\prime}}\mu_{j}(\alpha^{\prime})\left(  \mu_{k}(\alpha^{\prime
})\right)  ^{\ast}\langle\psi_{f^{-1}(\alpha^{\prime})}|\rho_{r}|\psi
_{f^{-1}(\alpha^{\prime})}\rangle d(\alpha^{\prime})\right)  . \label{max}%
\end{equation}
Our task is to maximize $\overline{F}(\mathcal{M},T)$ under the constraints A,
B, C and D. A general approach makes use of Lagrange multipliers. Unless
otherwise stated, we take the estimated state $\rho_{r}$ to be a pure state
$|\varphi_{r}\rangle\langle\varphi_{r}|$. Let $\{|\chi_{j}\rangle:j=1,...,M\}$
be an orthonormal basis of $\mathcal{L}(S)$. With respect to this basis, we
can express the estimated state $|\varphi_{r}\rangle$ as $|\varphi_{r}%
\rangle=\sum_{i=j}^{M}\chi_{jr}|\chi_{j}\rangle$, with $\sum_{j=1}^{M}%
|\chi_{jr}|^{2}=1$ for every $r\in\Lambda$. The variables considered are
$C_{r},\lambda_{ir},\chi_{jr}$, where $r\in\Lambda$, $i\in\{1,...,N\}$ and
$j\in\{1,...,M\}$. In this paper, instead of make use of Lagrange multipliers,
we adopt an algebraic approach.

\section{Estimation of Bloch vectors: previous works}

We consider here the simplest case of state estimation, that is the problem of
estimating the direction of a pure qubit. In this section we sketch some of
the related previous works.

\subsection{The Bloch sphere representation}

Any state of a quantum system described by a two dimensional Hilbert space
$\mathcal{H}\cong\mathbb{C}^{2}$ is called\emph{\ qubit}. Spin states of an
electron and polarization states of a photon are examples of qubits. Any
\emph{pure qubit} is a vector of $\mathcal{H}$. There is an one-to-one
correspondence between normalized pure qubits and unit vectors of the
Euclidean space $\mathbb{R}^{3}$. This correspondence (which also valid for
normalized mixed qubits) is called the \emph{Bloch sphere representation} of
qubits. In this representation, any pure qubit $|\psi(\theta,\phi)\rangle
=\cos\frac{\theta}{2}|0\rangle+e^{i\phi}\sin\frac{\theta}{2}|1\rangle$,
corresponds to a \emph{Bloch vector} $\widehat{\mathbf{n}}=(\sin\theta\cos
\phi,\sin\theta\sin\phi,\cos\theta)$, where $\theta\in\lbrack0,\pi]$ and
$\phi\in\lbrack0,2\pi)$. More generally, $|\psi(\theta,\phi)\rangle\langle
\psi(\theta,\phi)|=\frac{1}{2}(I+\widehat{\mathbf{n}}\cdot\widehat
{\mathbf{\sigma}})$, where $I$ is the $2\times2$ identity matrix and
$\widehat{\mathbf{\sigma}}$ is the vector with $x$-, $y$- and $z$-component as
the Pauli spin matrices $\sigma_{x}$, $\sigma_{y}$ and $\sigma_{z}$,
respectively. We write $|\psi(\theta,\phi)\rangle=|\widehat{\mathbf{n}}%
\rangle$. Here, $|0\rangle$ and $|1\rangle$ are the eigenstates of $\sigma
_{z}$ corresponding to the eigenvalues $1$ and $-1$, respectively. The state
$|\psi(\pi-\theta,\pi+\phi)\rangle=\sin\frac{\theta}{2}|0\rangle-e^{i\phi}%
\cos\frac{\theta}{2}|1\rangle$, corresponding to the Bloch vector
$-\widehat{\mathbf{n}}$, is orthogonal to $|\psi(\theta,\phi)\rangle$.

\subsection{Peres-Wootters}

Let $S=\{|\psi_{1}\rangle,|\psi_{2}\rangle,|\psi_{3}\rangle\}$, where
$|\psi_{1}\rangle=|0\rangle$, $|\psi_{2}\rangle=\frac{1}{2}|0\rangle
+\frac{\sqrt{3}}{2}|1\rangle$ and $|\psi_{3}\rangle=\frac{1}{2}|0\rangle
-\frac{\sqrt{3}}{2}|1\rangle$. Let $S_{2}=\{|\psi_{1}\rangle^{\otimes2}%
,|\psi_{2}\rangle^{\otimes2},|\psi_{3}\rangle^{\otimes2}\}$ and $s(\mathcal{M}%
,r,j)=|\langle\varphi_{r}|\psi_{j}\rangle|^{2}$. The state $|\varphi
_{r}\rangle$ is the estimated qubit corresponding to the $r$-th measurement
outcome of the general POVM $\mathcal{M}=\{\widehat{E}_{r}=C_{r}\cdot
P[\lambda_{1r}|00\rangle+\lambda_{2r}|01\rangle+\lambda_{3r}|10\rangle
+\lambda_{4r}|11\rangle]:r\in\Lambda\}$ satisfying the constraints A, B, C and
D. Peres and Wootters \cite{pereswootters91} gave numerical evidence that
measurements with entangled bases $\lambda_{1r}|00\rangle+\lambda
_{2r}|01\rangle+\lambda_{3r}|10\rangle+\lambda_{4r}|11\rangle$ can give rise
to higher average fidelity compared to the case when the measurement bases are
not entangled.%

\begin{center}
\includegraphics[
height=2.0783in,
width=2.3073in
]%
{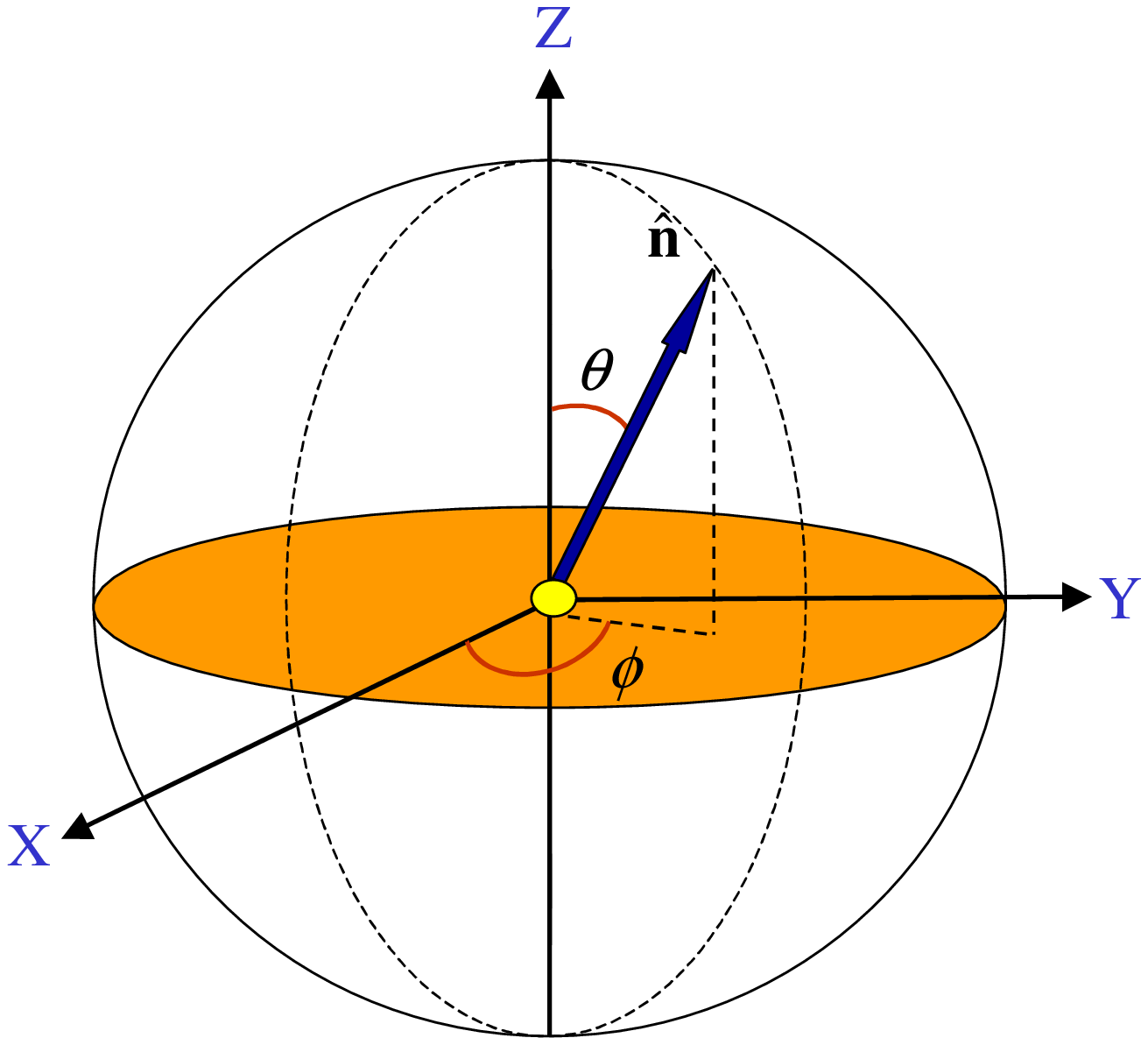}%
\\
Figure 1:
\end{center}

\subsection{Massar-Popescu\label{s}}

Massar and Popescu \cite{massarpopescu99} considered $S=\{|\psi(\theta
,\phi)\rangle:\theta\in\lbrack0,\pi],\phi\in\lbrack0,2\pi)\}$, $S_{n}%
=\{|\psi(\theta,\phi)\rangle^{\otimes n}:\theta\in\lbrack0,\pi],\phi\in
\lbrack0,2\pi)\}$ and $s(\mathcal{M},r,(\theta,\phi))=|\langle\varphi_{r}%
|\psi(\theta,\phi)\rangle|^{2}=\frac{1+\widehat{\mathbf{n}}\cdot
\widehat{\mathbf{n}}_{r}}{2}$, where $|\varphi_{r}\rangle=\cos\frac{\theta
_{r}}{2}|0\rangle+e^{i\phi_{r}}\sin\frac{\theta_{r}}{2}|1\rangle
=|\widehat{\mathbf{n}}_{r}\rangle$. The state $|\varphi_{r}\rangle$ is the
estimated qubit corresponding to the $r$-th measurement outcome of the POVM
$\mathcal{M}=\{\widehat{E}_{r}:r\in\Lambda\}$:

\begin{itemize}
\item If $n=1$ then $\Lambda=\{1,2\}$, $\widehat{E}_{1}=|0\rangle\langle0| $,
$\widehat{E}_{1}=|1\rangle\langle1|$, $|\varphi_{1}\rangle=|0\rangle$ and
$|\varphi_{2}\rangle=|1\rangle$. For $n=1$, $\overline{F}^{\max}=\frac{2}{3}.$

\item If $n=2$ then $\Lambda=\{1,2,3,4\}$ and $\widehat{E}_{j}=P[\frac{1}%
{2}|\psi^{-}\rangle+\frac{\sqrt{3}}{2}|\widehat{\mathbf{n}}_{j}\rangle
^{\otimes2}]$ for $j=1,...,4$, where $\widehat{\mathbf{n}}_{1}=(0,0,1)$,
$\widehat{\mathbf{n}}_{2}=\left(  \tfrac{\sqrt{8}}{3},0,-\frac{1}{3}\right)
$, $\widehat{\mathbf{n}}_{3}=\left(  \frac{-\sqrt{2}}{3},\sqrt{\frac{2}{3}%
},-\frac{1}{3}\right)  $ and $\widehat{\mathbf{n}}_{4}=\left(  \frac{-\sqrt
{2}}{3},-\sqrt{\frac{2}{3}},-\frac{1}{3}\right)  $; $|\psi^{-}\rangle=\frac
{1}{\sqrt{2}}(|01\rangle-|10\rangle)$ and $|\varphi_{j}\rangle=|\widehat
{\mathbf{n}}_{j}\rangle$ for $j=1,...,4$. For $n=2$, $\overline{F}^{\max
}=\frac{3}{4}$.

\item If $n>2$ then $\overline{F}^{\max}=\frac{n+1}{n+2}$, which was obtained
by making use of an infinite POVM (\emph{i.e.} for which $\Lambda$ is an
infinite set) known as \emph{covariant measurement} (see, \emph{e.g.},
\cite{h}).
\end{itemize}

\subsection{Derka-Buzek-Ekert}

For the general case considered by Massar and Popescu, Derka \emph{et al.}
\cite{derka98} have given a finite POVM such that $\overline{F}^{\max}%
=\frac{n+1}{n+2}$ for any $n$. In addition, they also considered
$S=\{|\psi(\tfrac{\pi}{2},\phi)\rangle:\phi\in\lbrack0,2\pi)\}$, with
$S_{n}=\{|\psi(\tfrac{\pi}{2},\phi)\rangle^{\otimes n}:\phi\in\lbrack
0,2\pi)\}$, $\Lambda=\{0,...,n\}$ and $\widehat{E}_{r}=P[\frac{1}{\sqrt{n+1}%
}\sum\limits_{j=0}^{n}e^{\frac{2\pi ij}{n+1}}|S_{j}^{(n)}\rangle]$, where%

\begin{equation}
|S_{j}^{(n)}\rangle=\frac{1}{\sqrt{\binom{n}{j}}}\sum
\limits_{\substack{_{\substack{x_{i}=0,1 \\1\leq i\leq n}} \\|\{x_{i}%
:x_{i}=0\}|=j}}|x_{1}x_{2}\cdots x_{n}\rangle\label{exp1}%
\end{equation}
The state $|S_{j}^{(n)}\rangle$ is the symmetrized $n$-qubit superposition of
$j$ $0$'s and $(n-j)$ $1$'s, $|\varphi_{r}\rangle=|\psi(\tfrac{\pi}{2}%
,\tfrac{2\pi r}{n+1})\rangle$ and
\[
\overline{F}^{\max}=\frac{1}{2}+\frac{1}{2^{n+1}}\sum\limits_{i=0}^{n-1}%
\sqrt{\binom{n}{i}\binom{n}{i+1}}.
\]

\subsection{Latorre-Pascual-Tarrach}

Latorre \emph{et al.} \cite{latorre98} considered the case of estimation of
qubits for which $S=\{|\psi(\theta,\phi)\rangle:\theta\in\lbrack0,\pi],\phi
\in\lbrack0,2\pi)\}$, $S_{n}=\{|\psi(\theta,\phi)\rangle^{\otimes n}:\theta
\in\lbrack0,\pi],\phi\in\lbrack0,2\pi)\}$ and $s(\mathcal{M},r,(\theta
,\phi))=\frac{1+\widehat{\mathbf{n}}\cdot\widehat{\mathbf{n}}_{r}}{2}$. The
value $\overline{F}^{\max}=\frac{n+1}{n+2}$ corresponds to the estimation
strategy which uses a POVM with elements $\widehat{E}_{r}=C_{r}P[|\psi
(\theta_{r},\phi_{r})\rangle^{\otimes n}]$, and $|\varphi_{r}\rangle
=|\psi(\theta_{r},\phi_{r})\rangle$. The table below contains the parameters
of the strategy for $2\leq n\leq5$. For $n>5$, the minimal finite POVM could
be established.
\begingroup
\scalefont{0.7}%
%

\[%
\begin{tabular}
[c]{c|c|c|c|c|c}%
$n$ & $r$ & $C_{r}$ & $\phi_{r}/\pi$ & $\cos\theta_{r}$ & $\overline{F}^{\max
}$\\\hline
$2$ &
\begin{tabular}
[c]{c}%
$1$\\
$2-4$%
\end{tabular}
& $3/4$ &
\begin{tabular}
[c]{c}%
$0$\\
$2(r-2)/3$%
\end{tabular}
&
\begin{tabular}
[c]{r}%
$1$\\
$-1/3$%
\end{tabular}
& $3/4$\\\hline
$3$ &
\begin{tabular}
[c]{c}%
$1$\\
$2$\\
$3-6$%
\end{tabular}
& $2/3$ &
\begin{tabular}
[c]{c}%
$0$\\
$0$\\
$(r-3)/2$%
\end{tabular}
& \multicolumn{1}{|r|}{%
\begin{tabular}
[c]{r}%
$1$\\
$-1$\\
$0$%
\end{tabular}
} & $4/5$\\\hline
$4$ &
\begin{tabular}
[c]{c}%
$1$\\
$2$\\
$3-6$\\
$7-10$%
\end{tabular}
&
\begin{tabular}
[c]{c}%
$5/12$\\
$5/12$\\
$25/48$\\
$25/48$%
\end{tabular}
&
\begin{tabular}
[c]{c}%
$0$\\
$0$\\
$(r-3)/2$\\
$\,(r-\frac{13}{2})/2$%
\end{tabular}
&
\begin{tabular}
[c]{r}%
$1$\\
$-1$\\
$1/\sqrt{5}$\\
$-1/\sqrt{5}$%
\end{tabular}
& $5/6$\\\hline
$5$ &
\begin{tabular}
[c]{c}%
$1$\\
$2$\\
$3-7$\\
$8-12$%
\end{tabular}
& $1/2$ &
\begin{tabular}
[c]{c}%
$0$\\
$0$\\
$2(r-3)/5$\\
$2(r-\frac{15}{2})/5$%
\end{tabular}
&
\begin{tabular}
[c]{r}%
$1$\\
$-1$\\
$1/\sqrt{5}$\\
$-1/\sqrt{5}$%
\end{tabular}
& $5/7$%
\end{tabular}
\]
%

\endgroup

\subsection{Gisin-Popescu-Massar}

Gisin and Popescu \cite{gisinpopescu99} considered the problem of estimating
the direction a qubit $|\psi(\theta,\phi)\rangle$ from the entire Bloch
sphere, when an \emph{anti-parallel} state $|\psi(\theta,\phi)\rangle
\otimes|\psi(\pi-\theta,\pi+\phi)\rangle$ is supplied with equal probability
over the set $(\theta,\phi)$. Then, let $S=\{|\psi(\theta,\phi)\rangle
:\theta\in\lbrack0,\pi],\phi\in\lbrack0,2\pi)\}$ and $S_{2}=\{|\psi
(\theta,\phi)\rangle\otimes|\psi(\pi-\theta,\pi+\phi)\rangle:\theta\in
\lbrack0,\pi],\phi\in\lbrack0,2\pi)\}$. The state $|\varphi_{r}\rangle
=|\widehat{\mathbf{n}}_{r}\rangle$ is the estimated qubit corresponding to the
$r$-th measurement outcome of the POVM $\mathcal{M}=\{\widehat{E}_{r}%
:r\in\Lambda\}$: $\widehat{E}_{r}=P[\alpha|\widehat{\mathbf{n}}_{r}%
,-\widehat{\mathbf{n}}_{r}\rangle-\beta\sum_{\substack{k=1 \\k\neq r}%
}^{n}|\widehat{\mathbf{n}}_{k},-\widehat{\mathbf{n}}_{k}\rangle], $ for
$r\in\{1,2,3,4\}=\Lambda$, where $\alpha=\frac{13}{6\sqrt{6}-2\sqrt{2}}$ and
$\beta=\tfrac{5-2\sqrt{3}}{6\sqrt{6}-2\sqrt{2}}$. The average fidelity for
this strategy was shown to be $\overline{F}(\mathcal{M},T)=\frac{5\sqrt{3}%
+33}{3\left(  3\sqrt{3}-1\right)  ^{2}}$. Massar \cite{massar00} established
that this strategy is optimal. Moreover, Massar proved that in order to
estimate the direction of a vector that lies on the plane perpendicular to the
direction of the Bloch vector of the qubit $|\psi(\theta,\phi)\rangle$ (so
$s(\mathcal{M},T,r,(\theta,\phi))=1-\left(  \widehat{\mathbf{n}}\cdot
\widehat{\mathbf{n}}_{r}\right)  ^{2}$), parallel and anti-parallel states
give the optimal fidelities $\overline{F}_{s}^{\max}=0.8$ and $\overline
{F}_{s}^{\max}=0.733$, respectively. Thus, parallel qubits give then better
fidelity for the chosen score.

\subsection{LOCC Measurements}

Let us suppose that we are restricted to perform measurements on individual
qubits and then use the measurements results for one qubit to perform
measurements on another qubit and so on and so forth (this procedure is known
as \emph{LOCC measurement}). For every encoding, the supplied multiqubit state
would then contain the same amount of information about the direction of the
qubit as far as encoding is done in terms of product states. One can ask: what
kind of measurement provides more information about the direction of the qubit
(LOCC or an entangled one)? Gill and Massar \cite{gillmassar00} have shown
that, as $n\rightarrow\infty$, the difference between optimal fidelities for
LOCC and entangled measurements (on $n$ copies of the qubit) goes to zero.
This is true not only for encodings of the form $|\psi\rangle^{\otimes n}$,
but also for any other kind of product state encoding. For similar results see
also Bagan \emph{et al.} \cite{bg}.

\section{Estimation of parallel and anti-parallel qubits}

We consider here the problem of estimating the direction of pure qubit taken
from a circle (for given $\theta$), when $n$ copies of the qubit are supplied
with equal probability. The qubit which we are going to estimate belongs then
to the set $S_{\theta}=\{|\psi(\theta,\phi)\rangle=\cos\frac{\theta}%
{2}|0\rangle+e^{i\phi}\sin\frac{\theta}{2}|1\rangle:\phi\in\lbrack0,2\pi)\}$.
The set of supplied states is $S_{n}^{(\theta)}=\{|\psi(\theta,\phi
)\rangle^{\otimes n}:\phi\in\lbrack0,2\pi)\}$, where $|\varphi_{r}\rangle
=\cos\frac{\theta}{2}|0\rangle+e^{i\phi_{r}}\sin\frac{\theta}{2}%
|1\rangle=|\widehat{\mathbf{n}}_{r}\rangle$. Figure 2 illustrates this
setting.
\begin{center}
\includegraphics[
height=2.081in,
width=2.31in
]%
{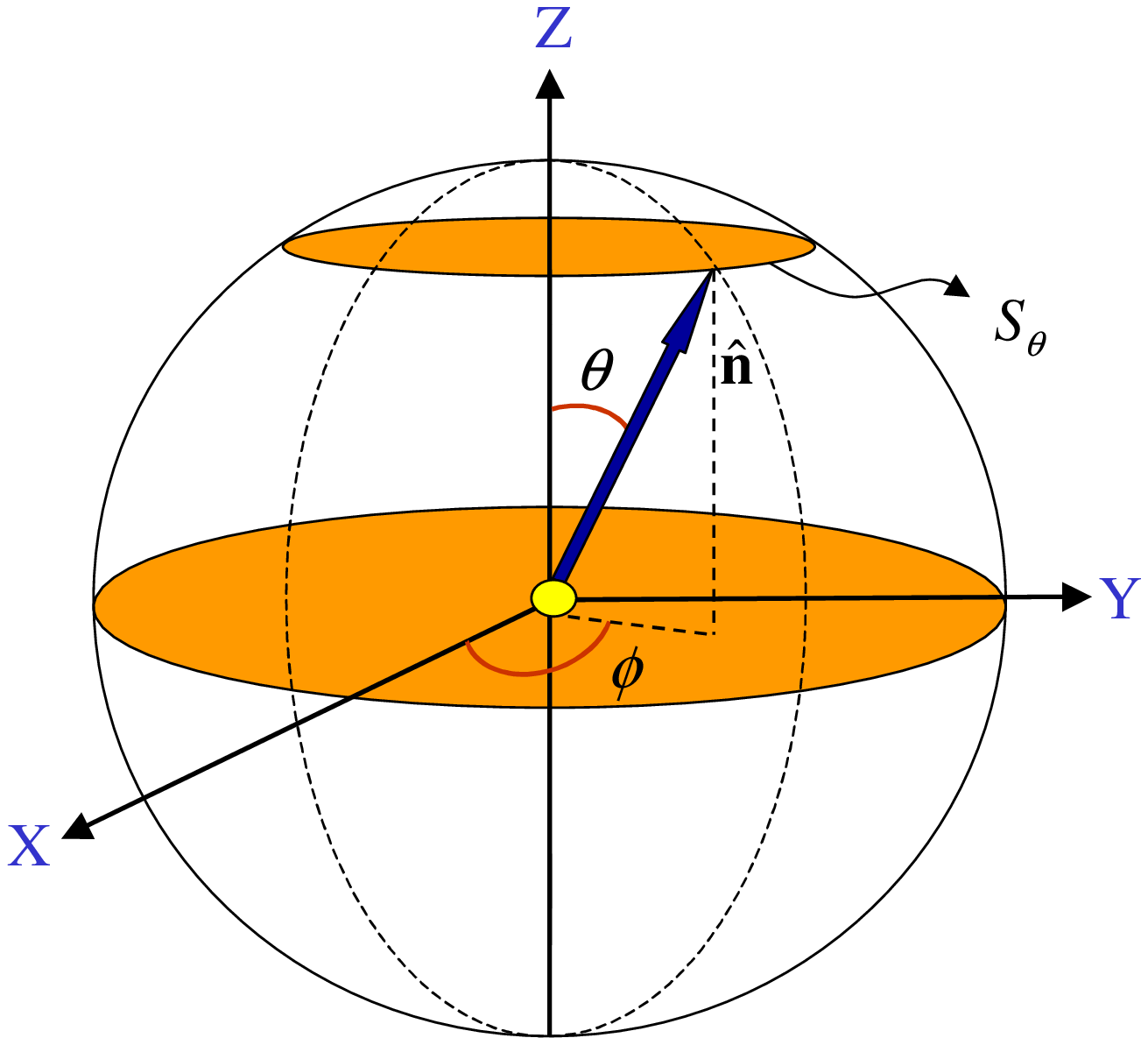}%
\\
Figure 2:
\label{circle}%
\end{center}

The score is $s(\mathcal{M},r,(\theta,\phi))=|\langle\varphi_{r}|\psi
(\theta,\phi)\rangle|^{2}=\tfrac{1+\widehat{\mathbf{n}}\cdot\widehat
{\mathbf{n}}_{r}}{2}$. The elements of $S_{n}^{(\theta)}$ can be written as
\[
|\psi(\theta,\phi)\rangle^{\otimes n}=\sum_{j=0}^{n}\sqrt{\binom{n}{j}}\left(
\cos\frac{\theta}{2}\right)  ^{j}\left(  \sin\frac{\theta}{2}\right)
^{n-j}e^{i\left(  n-j\right)  \phi}|S_{j}^{(n)}\rangle=|\Psi_{n,0}(\theta
,\phi)\rangle,
\]
where $|S_{j}^{(n)}\rangle$ is expressed in (\ref{exp1}). More generally, for
any fixed $\theta\in\lbrack0,\pi]$, for estimating the direction of
$|\psi(\theta,\phi)\rangle\in S_{\theta}$, we can consider the scenario in
which the state is supplied with equal probability from the set
\begin{equation}
S_{n,m}^{(\theta)}=\{|\psi(\theta,\phi)\rangle^{\otimes n}\otimes|\psi
(\pi-\theta,\pi+\phi)\rangle^{\otimes m}:\phi\in\lbrack0,2\pi)\}. \label{snm}%
\end{equation}
We use the notation
\[
|\Psi_{n,m}(\theta,\phi)\rangle=|\psi(\theta,\phi)\rangle^{\otimes n}%
\otimes|\psi(\pi-\theta,\pi+\phi)\rangle^{\otimes m}.
\]
Again, we take here $|\varphi_{r}\rangle=|\psi(\theta,\phi_{r})\rangle$. Any
state of $S_{n,m}^{(\theta)}$ can be then written as%
\[
|\Psi_{n,m}(\theta,\phi)\rangle=\sum_{p=0}^{n+m}e^{i\left(  n+m-p\right)
\phi}\mathcal{N}_{p}(\theta)|\xi_{p}(\theta)\rangle,
\]
where%
\[
|\xi_{p}(\theta)\rangle=\frac{1}{\mathcal{N}_{p}(\theta)}\sum_{\left(
k,l\right)  \in T_{p}}^{n+m}\sqrt{\binom{n}{k}\binom{m}{l}}\left(  \cos
\frac{\theta}{2}\right)  ^{2\left(  m-l+k\right)  }\left(  \sin\frac{\theta
}{2}\right)  ^{2\left(  n+l-k\right)  }\left(  -1\right)  ^{m-l}|S_{k}%
^{(n)}\rangle\otimes|S_{l}^{(m)}\rangle,
\]%
\[
\mathcal{N}_{p}(\theta)=\left(  \sum_{\left(  k,l\right)  \in T_{p}}%
^{n+m}\binom{n}{k}\binom{m}{l}\left(  \cos\frac{\theta}{2}\right)  ^{2\left(
m-l+k\right)  }\left(  \sin\frac{\theta}{2}\right)  ^{2\left(  n+l-k\right)
}\right)  ^{\frac{1}{2}}%
\]
and
\[
T_{p}=\{(k,l)\in\{0,...,n\}\times\{0,...,m\}:k+l=p\}.
\]
Following the description given in Section 3, the elements of the most general
POVM, which may appear in an estimation strategy are of the form
\[%
\begin{tabular}
[c]{lll}%
$\widehat{E}_{r}=C_{r}^{(\theta)}P\left[  \sum_{p=0}^{n+m}\lambda_{rp}%
(\theta)|\xi_{p}(\theta)\rangle\right]  ,$ &  & $\text{for every }r\in
\Lambda,$%
\end{tabular}
\]
where $C_{r}^{(\theta)}>0$, $\sum_{p=0}^{n+m}|\lambda_{rp}(\theta)|^{2}=1$ for
every $r\in\Lambda$, and%
\begin{equation}%
\begin{tabular}
[c]{lll}%
$\sum_{r\in\Lambda}C_{r}^{(\theta)}\lambda_{rp}(\theta)(\lambda_{rq}%
(\theta))^{\ast}=\delta_{pq},$ &  & for all $p,q\in\{0,1,....,n+m\}$.
\end{tabular}
\label{con1}%
\end{equation}
The average fidelity corresponding to this estimation strategy will be denoted
by $\overline{F}_{n,m}(\theta)$. Using the POVM described in the previous
section, we obtain%

\[
\overline{F}_{n,m}(\theta)=\frac{1+\cos^{2}\theta}{2}+\frac{\sin^{2}\theta}%
{2}\sum_{p=0}^{n+m}\mathcal{N}_{p-1}(\theta)\mathcal{N}_{p}(\theta)\sum
_{r\in\Lambda}C_{r}^{(\theta)}\operatorname{Re}\left(  \lambda_{r\left(
p-1\right)  }(\theta)(\lambda_{rp}(\theta))^{\ast}e^{-i\phi_{r}}\right)
\]
and we observe that%
\begin{equation}
\overline{F}_{n,m}(\theta)\leq\frac{1+\cos^{2}\theta}{2}+\frac{\sin^{2}\theta
}{2}\sum_{p=0}^{n+m}\mathcal{N}_{p-1}(\theta)\mathcal{N}_{p}(\theta)\sum
_{r\in\Lambda}C_{r}^{(\theta)}|\lambda_{r\left(  p-1\right)  }(\theta
)\times\lambda_{rp}(\theta)|. \label{step1}%
\end{equation}
Then by the Schwartz inequality,%
\begin{align}
\overline{F}_{n,m}(\theta)  &  \leq\frac{1+\cos^{2}\theta}{2}+\frac{\sin
^{2}\theta}{2}\sum_{p=1}^{n+m}\mathcal{N}_{p-1}(\theta)\mathcal{N}_{p}%
(\theta)\left[  \sum_{r\in\Lambda}C_{r}^{(\theta)}|\lambda_{r\left(
p-1\right)  }(\theta)|^{2}\right]  ^{\frac{1}{2}}\left[  \sum_{r\in\Lambda
}C_{r}^{(\theta)}|\lambda_{rp}(\theta)|^{2}\right]  ^{\frac{1}{2}%
}\label{step2}\\
&  =\frac{1+\cos^{2}\theta}{2}+\frac{\sin^{2}\theta}{2}\sum_{p=1}%
^{n+m}\mathcal{N}_{p-1}(\theta)\mathcal{N}_{p}(\theta),\nonumber
\end{align}
which follows from (6). We then see that
\[
\overline{F}_{n,m}^{\max}(\theta)\leq\frac{1+\cos^{2}\theta}{2}+\frac{\sin
^{2}\theta}{2}\sum_{p=1}^{n+m}\mathcal{N}_{p-1}(\theta)\mathcal{N}_{p}%
(\theta),
\]
which is an upper bound on $\overline{F}_{n,m}^{\max}(\theta)$ independent of
any measurement strategy. We describe now an estimation strategy which attains
this quantity. Equality in (\ref{step2}) holds if and only if
\[%
\begin{tabular}
[c]{lll}%
$C_{r}^{(\theta)}|\lambda_{r\left(  p-1\right)  }(\theta)|^{2}=K_{p}%
C_{r}^{(\theta)}|\lambda_{rp}(\theta)|^{2},$ &  & for every $r\in\Lambda$,
\end{tabular}
\]
where $K_{p}$ is constant for $p=0,1,...,n+m$. It follows from the condition
(\ref{con1}) that $K_{p}=1$ for every $p=0,1,...,n+m$. This implies that
\begin{equation}%
\begin{tabular}
[c]{lll}%
$\lambda_{rp}(\theta)=\frac{e^{i\varepsilon_{rp}}}{\sqrt{n+m+1}},$ &  &
$\text{where }\varepsilon_{rp}\in\mathbb{R}\text{ for every }%
p=0,1,...,n+m\text{ and }r\in\Lambda.$%
\end{tabular}
\label{anyone}%
\end{equation}
Using (\ref{anyone}), we see that equality in (\ref{step1}) holds if and only
if $\varepsilon_{rp}=2n_{rp}\pi+\varepsilon_{r(p+1)}+\phi_{r}$, for each
$r\in\Lambda$ and each $p=0,1,...,n+m$, where $n_{rp}\in\mathbb{Z}$. Then
\begin{equation}%
\begin{tabular}
[c]{ll}%
$\varepsilon_{rp}=2L_{rp}\pi+\varepsilon_{r(n+m)}+(n+m-p)\phi_{r},$ &
$\text{where }L_{rp}\in\mathbb{Z}\text{ for every }r\in\Lambda\text{ and
}p=0,1,...,n+m.$%
\end{tabular}
\label{anyone2}%
\end{equation}
Using (\ref{anyone}) and (\ref{anyone2}) into (\ref{con1}), we can write%
\begin{equation}%
\begin{tabular}
[c]{lll}%
$\sum_{r\in\Lambda}C_{r}^{(\theta)}e^{i\left(  q-p\right)  \phi_{r}%
}=(m+n+1)\delta_{pq},$ &  & for every $p,q=0,1,...,n+m.$%
\end{tabular}
\label{anyone3}%
\end{equation}
Thus, we see that one possible situation where the condition (\ref{anyone3})
is satisfied is given by $\Lambda=\{0,1,...,n+m\}$, $C_{r}^{(\theta)}=1$ for
all $r\in\Lambda$, and $\phi_{r}=\frac{2\pi r}{n+m+1}$ for all $r\in\Lambda$.
Taking these parameters in the estimation strategy, we define a POVM
$\mathcal{M}=\{\widehat{E}_{r}:r\in\Lambda\}$ such that%
\[
\widehat{E}_{r}=C_{r}^{(\theta)}P\left[  \sum_{p=0}^{n+m}\lambda_{rp}%
(\theta)|\xi_{p}(\theta)\rangle\right]  =P\left[  \frac{1}{\sqrt{m+m+1}}%
\sum_{p=0}^{n+m}\exp\left[  \frac{2\pi i\left(  n-m-p\right)  r}%
{n+m+1}\right]  |\xi_{p}(\theta)\rangle\right]  .
\]
Then%
\[
\overline{F}_{n,m}(\theta)=\frac{1+\cos^{2}\theta}{2}+\frac{\sin^{2}\theta}%
{2}\sum_{p=1}^{n+m}\mathcal{N}_{p-1}(\theta)\mathcal{N}_{p}(\theta
)=\overline{F}_{n,m}^{\max}(\theta).
\]
Note that the basis of the POVM is the Fourier basis of dimension $n+m+1$.

\subsection{The dimensional argument}

Gisin and Popescu \cite{gisinpopescu99} have shown that the anti-parallel
qubits $|\Psi_{1,1}(\theta,\phi)\rangle$ contain more information on an
average compared to parallel qubits $|\Psi_{2,0}(\theta,\phi)\rangle$,
regarding the direction of the qubit $|\psi(\theta,\phi)\rangle$, when
$\left(  \theta,\phi\right)  $ is uniformly distributed over $[0,\pi
]\times\lbrack0,2\pi)$. This is counterintuitive according to the reasoning in
classical physics. In fact, in order to get information about the direction of
a classical vector $\mathbf{v}$, either we can consider the parallel vectors
$\{\mathbf{v,v\}}$ or the anti-parallel vectors $\{\mathbf{v},-\mathbf{v}\}$
(when we are restricted to only to these two types of vectors). The parallel
and the anti-parallel vectors do not make any difference in this regard. This
is simply because $\mathbf{v}$ and $-\mathbf{v}$ contains the same information
about the direction of $\mathbf{v}$. However, a notable property of parallel
and anti-parallel qubits is the following: $\dim\mathcal{L}(\{|\Psi
_{1,1}(\theta,\phi)\rangle:\theta\in\lbrack0,\pi],\phi\in\lbrack0,2\pi)\})=4$
and $\dim\mathcal{L}(\{|\Psi_{2,0}(\theta,\phi)\rangle:\theta\in\lbrack
0,\pi],\phi\in\lbrack0,2\pi)\})=3$. Gisin and Popescu proposed the difference
in dimension as the reason behind the difference in optimal fidelities. We
call this reasoning as \emph{dimensional argument}. They support the
dimensional argument as follows. Even though
\[%
\begin{tabular}
[c]{lll}%
$|\langle\Psi_{1,1}(\theta,\phi)|\Psi_{1,1}(\theta^{\prime},\phi^{\prime
})\rangle|=|\langle\Psi_{2,0}(\theta,\phi)|\Psi_{2,0}(\theta^{\prime}%
,\phi^{\prime})\rangle|$, &  & where $\theta,\theta^{\prime}\in\lbrack0,\pi]$
and $\phi,\phi^{\prime}\in\lbrack0,2\pi)$,
\end{tabular}
\]
anti-parallel states are, as a \emph{whole}, \emph{farther apart} than
parallel states, because of the difference on the dimensions of the linear
spans. Note that,
\begin{align*}
|\langle\Psi_{1,1}(\theta,\phi)|\Psi_{1,1}(\theta^{\prime},\phi^{\prime
})\rangle|^{2}  &  =|\langle\psi(\theta,\phi)|\psi(\theta^{\prime}%
,\phi^{\prime})\rangle|^{2}\cdot|\langle\psi(\pi-\theta,\pi+\phi)|\psi
(\pi-\theta^{\prime},\pi+\phi^{\prime})\rangle|^{2}\\
&  =\left(  \frac{1+\widehat{\mathbf{n}}\cdot\widehat{\mathbf{n}}^{\prime}}%
{2}\right)  \cdot\left(  \frac{1+\left(  -\widehat{\mathbf{n}}\right)
\cdot(-\widehat{\mathbf{n}}^{\prime})}{2}\right)  =\left(  \frac
{1+\widehat{\mathbf{n}}\cdot\widehat{\mathbf{n}}^{\prime}}{2}\right)  ^{2}\\
&  =|\langle\Psi_{2,0}(\theta,\phi)|\Psi_{2,0}(\theta^{\prime},\phi^{\prime
})\rangle|^{2},
\end{align*}
where $|\psi(\theta,\phi)\rangle=|\widehat{\mathbf{n}}\rangle$ and
$|\psi(\theta^{\prime},\phi^{\prime})\rangle=|\widehat{\mathbf{n}}^{\prime
}\rangle$. Figure 3 clarifies the meaning of \emph{farther apart}. The three
unit vectors $\widehat{\mathbf{n}}_{1}$, $\widehat{\mathbf{n}}_{2}$ and
$\widehat{\mathbf{n}}_{3}$ lie on the equatorial plane and are linearly
dependent. The angle between each pair of them is $\alpha=\frac{2}{3}\pi$. We
consider now three linearly independent vectors $\widehat{\mathbf{m}}_{1}$,
$\widehat{\mathbf{m}}_{2}$ and $\widehat{\mathbf{m}}_{3}$, whose heads are on
a small circle such that the great circle joining north pole and the head of
$\widehat{\mathbf{n}}_{i}$ crosses the equator in the head of $\widehat
{\mathbf{m}}_{i}$. This means that the angle between $\widehat{\mathbf{m}}_{i}
$ and $\widehat{\mathbf{m}}_{j}$ is smaller than the angle $\alpha$ between
$\widehat{\mathbf{n}}_{i}$ and $\widehat{\mathbf{n}}_{j}$. In order to make
the angle between $\widehat{\mathbf{m}}_{i}$ and $\widehat{\mathbf{m}}_{j}$ to
be equal to $\alpha$ we need to rotate them in such a way that the distance
between their heads increases.
\begin{center}
\includegraphics[
height=2.081in,
width=2.3556in
]%
{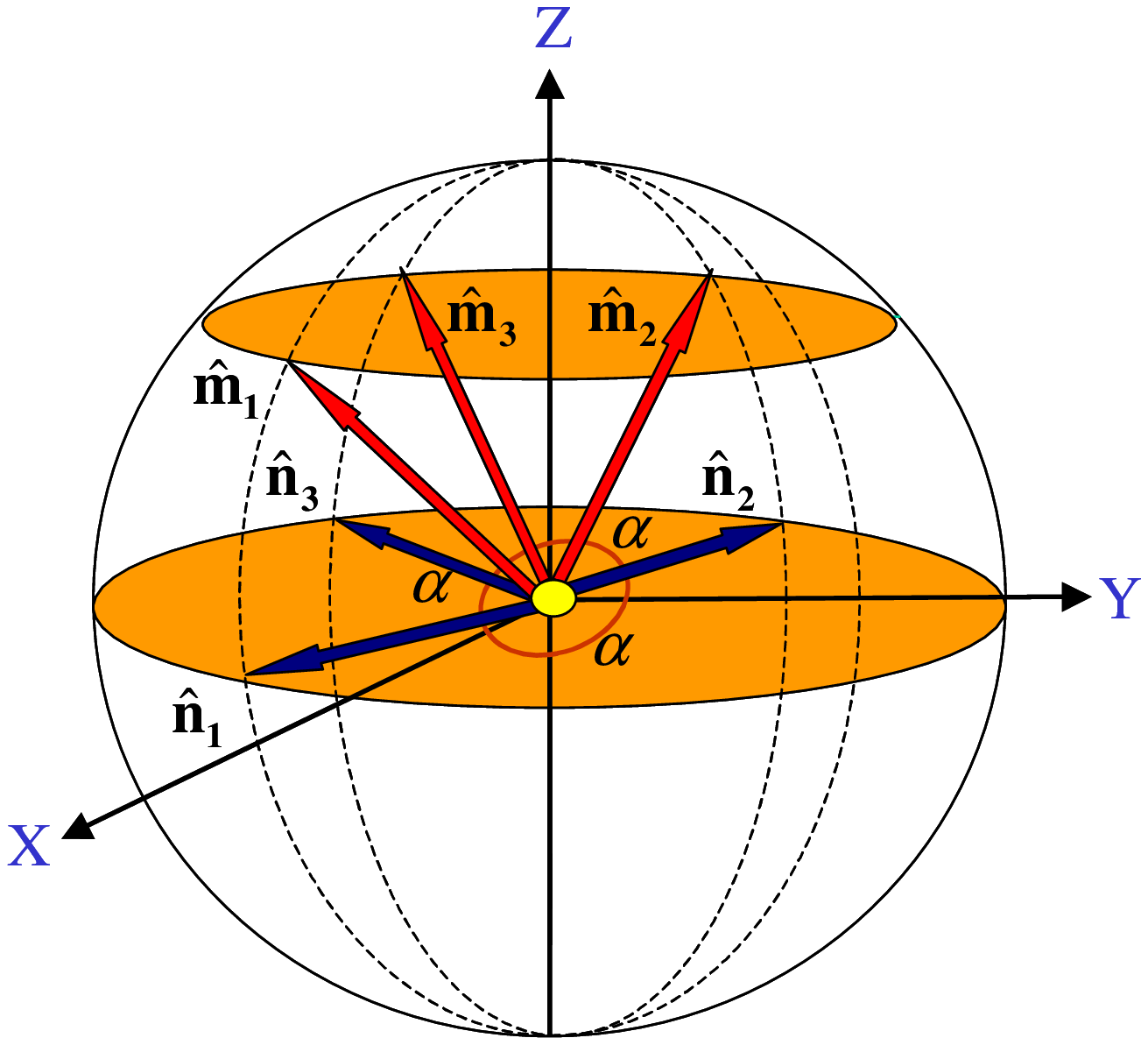}%
\\
Figure 3:
\label{3vect}%
\end{center}

The mathematical formulation distilled from the above argument can be
described as follows:

\bigskip

\noindent\textbf{Problem.} \emph{Let }$\mathcal{H}$\emph{\ be a }%
$d$\emph{-dimensional Hilbert space. Let }$\Delta$\emph{\ be a set of indices
(not necessarily finite). Let }$\mathbb{E}_{|\Phi\rangle}=\{(|\Phi_{i}%
\rangle,p_{i}:|\Phi_{i}\rangle\in\mathcal{H},0\leq p_{i}\leq1$\emph{\ for
every }$i\in\Delta$\emph{\ and }$\sum_{i\in\Delta}p_{i}=1\}$\emph{\ and}
$\mathbb{E}_{|\Upsilon\rangle}=\{\left(  |\Upsilon_{i}\rangle,p_{i}\right)
:|\Upsilon_{i}\rangle\in\mathcal{H},0\leq p_{i}\leq1$ \emph{for every }%
$i\in\Delta$\emph{\ and }$\sum_{i\in\Delta}p_{i}=1\}$\emph{. Suppose that
}$|\langle\Phi_{i}|\Phi_{j}\rangle|=|\langle\Upsilon_{i}|\Upsilon_{j}\rangle
|$\emph{\ for every }$i,j\in\Delta$\emph{, and that }$\dim\mathcal{L}%
(\{|\Phi_{i}\rangle:i\in\Delta\})>\dim\mathcal{L}(\{|\Upsilon_{i}\rangle
:i\in\Delta\})$\emph{. Then }$\overline{F}^{\max} $\emph{\ for estimating }%
$i$\emph{, for states given from }$\mathbb{E}_{|\Phi\rangle}$\emph{\ should be
grater than that for states given from }$\mathbb{E}_{|\Upsilon\rangle}$.

\bigskip

The problem can be restated in a more concrete form as follows. Let
$s:\Delta\times\Lambda\longrightarrow\lbrack0,1]$ be the score when a POVM
$\mathcal{M}=\{E_{r}:r\in\Lambda\}$ is applied on the unknown state $|\Phi
_{i}\rangle$, given from the set $\mathbb{E}_{|\Phi\rangle}$, with probability
$p_{i}$, and the $r$-th outcome has occurred. Note that the set of values of
the score $s$ are different for different forms of $\Lambda$ (\emph{i.e.} for
different choices of the POVM $\mathcal{M}$). Then the average fidelity is
\[
\overline{F}(\mathcal{M},\mathbb{E}_{|\Phi\rangle},s)=\sum_{r\in\Lambda}%
\sum_{i\in\Delta}p_{i}\langle\Phi_{i}|E_{r}|\Phi_{i}\rangle s(i,r).
\]
Similarly, for $\mathbb{E}_{|\Upsilon\rangle}$, we have $\overline
{F}(\mathcal{M},\mathbb{E}_{|\Upsilon\rangle},s)=\sum_{r\in\Lambda}\sum
_{i\in\Delta}p_{i}\langle\Upsilon_{i}|E_{r}|\Upsilon_{i}\rangle s(i,r)$. Let
$\overline{F}^{\max}(\mathbb{E}_{|\Phi\rangle},s)$ be the maximum of
$\overline{F}(\mathcal{M},\mathbb{E}_{|\Phi\rangle},s)$ over all possible
choices of the POVM $\mathcal{M}$ ($\overline{F}^{\max}(\mathbb{E}%
_{|\Upsilon\rangle},s)$ is defined similarly). Suppose that the following
conditions are satisfied simultaneously:

\begin{itemize}
\item $|\langle\Phi_{i}|\Phi_{j}\rangle|=|\langle\Upsilon_{i}|\Upsilon
_{j}\rangle|$, for every $i,j\in\Delta$;

\item $\dim\mathcal{L}(\{|\Phi_{i}\rangle:i\in\Delta\})>\dim\mathcal{L}%
(\{|\Upsilon_{i}\rangle:i\in\Delta\})$;

\item $\mathcal{L}(\{|\Phi_{i}\rangle:i\in\Delta\})\supset\mathcal{L}%
(\{|\Upsilon_{i}\rangle:i\in\Delta\})$.
\end{itemize}

Then we need to prove that $\overline{F}^{\max}(\mathbb{E}_{|\Phi\rangle
},s)>\overline{F}^{\max}(\mathbb{E}_{|\Upsilon\rangle},s)$. A solution to this
problem is still missing. So we do not know whether the statement of the above
problem can be taken as general principle.

\subsubsection{Entropic argument}

Consider the average density matrices
\[%
\begin{tabular}
[c]{lll}%
$\overline{\rho}_{2,0}=\frac{1}{4\pi}\int\limits_{\theta=0}^{\pi}%
\int\limits_{\phi=0}^{2\pi}P[|\Psi_{2,0}(\theta,\phi)\rangle]\sin\theta
d\theta d\phi$ & and & $\overline{\rho}_{1,1}=\frac{1}{4\pi}\int
\limits_{\theta=0}^{\pi}\int\limits_{\phi=0}^{2\pi}P[|\Psi_{1,1}(\theta
,\phi)\rangle]\sin\theta d\theta d\phi,$%
\end{tabular}
\]
associated to the ensembles for parallel and anti-parallel states%
\[%
\begin{tabular}
[c]{lll}%
$S_{2,0}^{(\theta)}=\{|\Psi_{2,0}(\theta,\phi)\rangle:\theta\in\lbrack
0,\pi],\phi\in\lbrack0,2\pi)\}$ & and & $S_{1,1}^{(\theta)}=\{|\Psi
_{1,1}(\theta,\phi)\rangle:\theta\in\lbrack0,\pi],\phi\in\lbrack0,2\pi)\}.$%
\end{tabular}
\]
Let $S(\rho)$ the \emph{von Neumann entropy} of a density matrix $\rho$. This
is defined as $S(\rho)=-\sum_{i}\lambda_{i}\log_{2}\lambda_{i}$, where
$\lambda_{i}$ is the $i$-th eigenvalue of $\rho$. The von Neumann entropy is a
measure of the information content of a density matrix. One can check that
$S(\overline{\rho}_{1,1})>S(\overline{\rho}_{2,0})$, therefore anti-parallel
states can be better distinguished, and hence, they posses more information
about the qubit. It should be noted that, even if the qubits $|\psi
(\theta,\phi)\rangle$ belong to the circle $S_{\theta}$, the von Neumann
entropy $S(\overline{\rho}_{1,1}(\theta))$ of the average density matrix of
the supplied anti-parallel qubits is greater than or equal to the von Neumann
entropy $S(\overline{\rho}_{2,0}(\theta))$ of the average density matrix of
the supplied parallel qubits (see Figure 5, right). This argument does not
hold in general. We provide a conterexample. Consider the following two
ensembles%
\[%
\begin{tabular}
[c]{lll}%
$\mathcal{E}_{1}=\{|0\rangle,\frac{1}{2}+\frac{\sqrt{2}+1/2}{4};|1\rangle
,\frac{1}{2}-\frac{\sqrt{2}+1/2}{4}\}$ & and & $\mathcal{E}_{2}=\{|0\rangle
,\frac{1}{2};\frac{1}{\sqrt{2}}\left(  |0\rangle+|1\rangle\right)  ,\frac
{1}{2}\}.$%
\end{tabular}
\]
Then%
\[%
\begin{tabular}
[c]{l}%
$S\left(  \frac{1}{2}P[|0\rangle]+\frac{1}{2}P\left[  \frac{1}{\sqrt{2}%
}\left(  |0\rangle+|1\rangle\right)  \right]  \right)  =H\left(  \frac{1}%
{2}+\frac{\sqrt{2}}{4}\right)  $\\
$>S\left(  \left(  \frac{1}{2}+\frac{\sqrt{2}+1/2}{4}\right)  P[|0\rangle
]+\left(  \frac{1}{2}-\frac{\sqrt{2}+1/2}{4}\right)  P[|1\rangle]\right)
=H\left(  \frac{1}{2}-\frac{\sqrt{2}+1/2}{4}\right)  ,$%
\end{tabular}
\]
where
\[%
\begin{tabular}
[c]{lll}%
$H(x)=-x\log_{2}x-\left(  1-x\right)  \log_{2}\left(  1-x\right)  $ & for &
$0\leq x\leq1.$%
\end{tabular}
\]
This shows that, even though the states in the ensemble $\mathcal{E}_{1}$ are
better distinguished than the states in $\mathcal{E}_{2}$ (as the states in
$\mathcal{E}_{1}$ are orthogonal to each other and the states in
$\mathcal{E}_{2}$ are not), the information content of the density matrix
corresponding to $\mathcal{E}_{1}$ is less than that the one corresponding to
$\mathcal{E}_{2}$.

\subsection{Inadequacy of the dimensional argument}

First of all, observe that, for any fixed $\theta\in(0,\pi)$, we have seen
that so far as $(n+m)$ is fixed, the $\mathcal{L}(S_{n,m}^{(\theta)})$ is
$(n+m+1)$-dimensional subspace of the $2^{n+m}$-dimensional Hilbert space of
the system. Of course the subspace $\mathcal{L}(S_{n,m}^{(\theta)})$ is
different for different values of $n$ and $m$. Thus, if $\overline{F}%
_{n,m}^{\max}(\theta)\neq\overline{F}_{n^{\prime},m^{\prime}}^{\max}(\theta)$,
for $n+m=n^{\prime}+m^{\prime}$, the dimensionality argument cannot be used to
explain this difference. We provide here three cases: (1) Let $n+m=2$. Then
$(n,m)$ is either $(2,0),(1,1)$ or $(0,2)$. (2) Let $n+m=3$. Then $(n,m)$ is
either $(3,0),(2,1),(1,2)$ or $(0,3)$. (3) Let $n+m=4$. Then
$(4,0),(3,1),(2,2),(1,3)$ or $(0,4)$. Note that $\overline{F}_{n,m}%
(\theta)\neq\overline{F}_{m,n}(\theta)$, since by swapping we can obtain the
state $|\psi(\theta,\phi)\rangle^{\otimes n}\otimes|\psi^{\perp}(\theta
,\phi)\rangle^{\otimes m}$ from $|\psi^{\perp}(\theta,\phi)\rangle^{\otimes
n}\otimes|\psi(\theta,\phi)\rangle^{\otimes m}$. These cases are illustrate by
Figure 4 ((1) left and (2) right) and Figure 5 ((3) left). In none of these
three figures the minimum of $\overline{F}_{n,0}^{\max}(\theta)$ is attained
at $\theta=\pi/2$. This is attained at two points symmetrically about $\pi/2$.
This phenomenon is somehow unexpected. As the circle $S_{\theta}$ is going far
and far from the poles towards the equator, we loose more and more information
about the direction of $|\psi\rangle$ (in $S_{\theta}$). It is then expected
that the optimal fidelity for states in $S_{\theta}$ (when the supplied state
is of the form $|\psi(\theta,\phi\rangle^{\otimes n}\otimes|\psi(\pi
-\theta,\pi+\phi)\rangle^{\otimes m}$) would start to decrease from $\theta
=0$, attaining its minimum at $\theta=\pi/2$, and again start to increase,
attaining its maximum at $\theta=\pi$.%

\begin{center}
\includegraphics[
height=2.2416in,
width=5.0999in
]%
{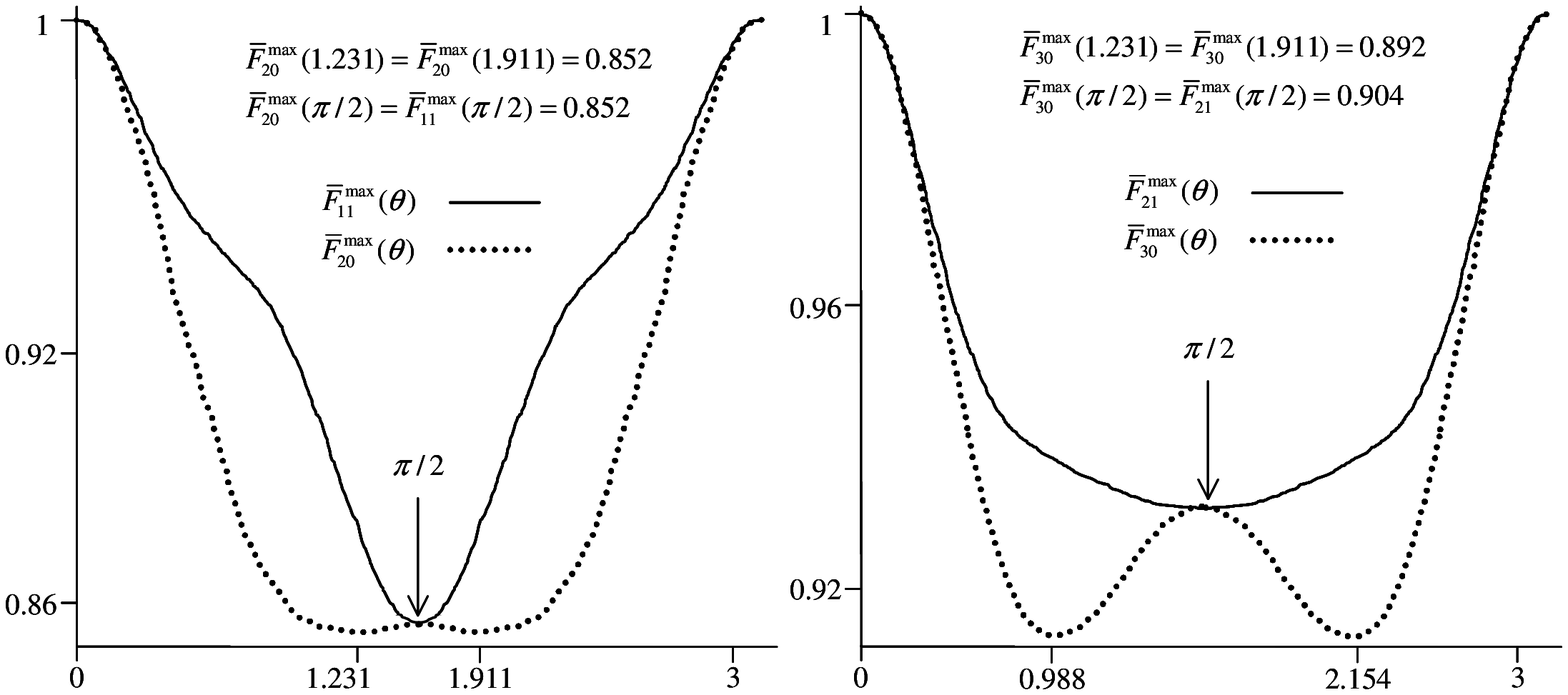}%
\\
Figure 4:
\label{12curve}%
\end{center}
%

\begin{center}
\includegraphics[
height=2.2489in,
width=5.0972in
]%
{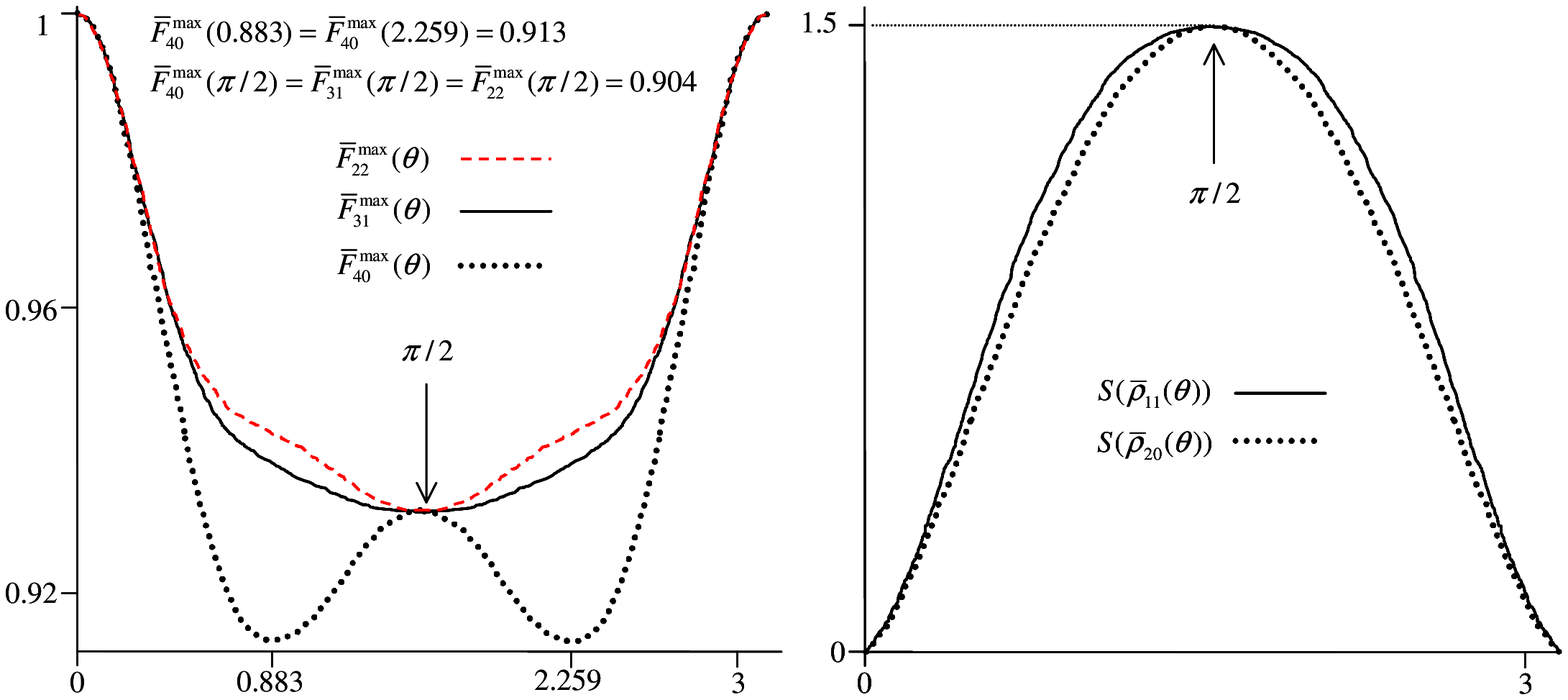}%
\\
Figure 5:
\label{23curve}%
\end{center}

\section{Estimation of qubits from two circles}

Consider the problem of estimating the direction of the Bloch vector
${\hat{\mathbf{n}}}=(\mathrm{sin}\theta~\mathrm{cos}\phi,~\mathrm{sin}%
\theta~\mathrm{sin}\phi,~\mathrm{sin}\theta)$ of a qubit $|\psi(\theta
,\phi)\rangle=~\mathrm{cos}\frac{\theta}{2}|0\rangle+e^{i\phi}~\mathrm{sin}%
\frac{\theta}{2}|1\rangle$, contained within the circle $S_{\theta}%
=\{|\psi(\theta,\phi)\rangle:\phi\in\lbrack0,2\pi)\}$, where $\theta\in
\lbrack0,\pi]$ is arbitrary but fixed. We have seen that in the case of
estimating the direction of the Bloch vector of the qubit $|\psi(\theta
,\phi)\rangle\in S_{\theta}$, the anti-parallel qubits $\left\vert {\Psi
}_{1,1}\right\rangle =|\psi(\theta,\phi)\rangle\otimes|\psi(\pi-\theta
,\pi+\phi)\rangle$ give better information compared to the parallel qubits
$\left\vert {\Psi}_{2,0}\right\rangle =|\psi(\theta,\phi)\rangle\otimes
|\psi(\theta,\phi)\rangle$ (where $\phi$ is uniformly distributed over
$[0.2\pi)$). A generalization of these two kind of encodings of the initial
two qubits is of the following form. The supplied two-qubit state is of the
form
\begin{equation}
\left\vert \Psi\left(  \theta,{\theta}_{0},\phi\right)  \right\rangle
=|\psi(\theta,\phi)\rangle\otimes\left\vert \psi\left(  \theta+{\theta}%
_{0},\phi\right)  \right\rangle , \label{eq1}%
\end{equation}
where ${\theta}_{0}$ is an arbitrary but fixed element from the set
$[-\theta,\pi-\theta]$, and $\phi$ is uniformly distributed over $[0,2\pi)$
(see Figure 6).%
\begin{center}
\includegraphics[
height=2.081in,
width=2.5007in
]%
{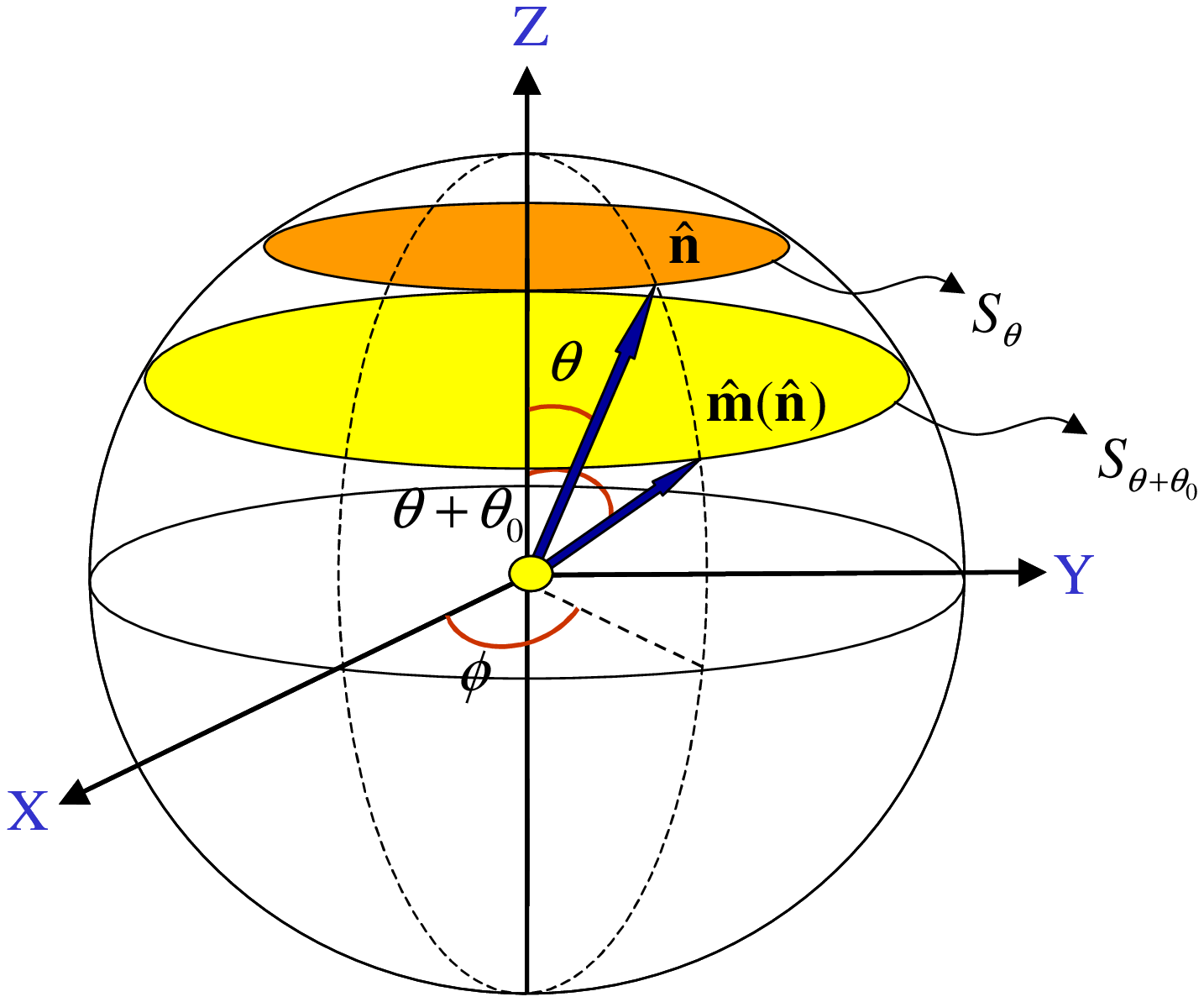}%
\\
Figure 6:
\label{2cir}%
\end{center}

One can check that
\begin{equation}
\left\vert \Psi\left(  \theta,{\theta}_{0},\phi\right)  \right\rangle
=~\mathrm{cos}\frac{\theta}{2}~\mathrm{cos}\frac{\theta+{\theta}_{0}}%
{2}|00\rangle+e^{2i\phi}~\mathrm{sin}\frac{\theta}{2}~\mathrm{sin}\frac
{\theta+{\theta}_{0}}{2}|11\rangle+e^{i\phi}N\left(  \theta,{\theta}%
_{0}\right)  \left\vert \chi\left(  \theta,{\theta}_{0}\right)  \right\rangle
, \label{eq2}%
\end{equation}
where
\begin{equation}%
\begin{array}
[c]{lcl}%
\left\vert \chi\left(  \theta,{\theta}_{0}\right)  \right\rangle  & = &
\displaystyle{\frac{1}{N\left(  \theta,{\theta}_{0}\right)  }\left(
\mathrm{cos}\frac{\theta}{2}~\mathrm{sin}\frac{\theta+{\theta}_{0}}%
{2}|01\rangle+~\mathrm{sin}\frac{\theta}{2}~\mathrm{cos}\frac{\theta+{\theta
}_{0}}{2}|10\rangle\right)  },\\
&  & \\
N\left(  \theta,{\theta}_{0}\right)  & = & \displaystyle\sqrt{\mathrm{cos}%
^{2}\frac{\theta}{2}~\mathrm{sin}^{2}\frac{\theta+{\theta}_{0}}{2}%
+~\mathrm{sin}^{2}\frac{\theta}{2}~\mathrm{cos}^{2}\frac{\theta+{\theta}_{0}%
}{2}}.
\end{array}
\label{eq3}%
\end{equation}
Thus, varying $\phi$ over $[0,2\pi)$, we see that for fixed $\theta$ and
${\theta}_{0}$, the set (which we denote as $S_{1,1}\left(  \theta,{\theta
}_{0}\right)  $) of all states $\left\vert \Psi\left(  \theta,{\theta}%
_{0},\phi\right)  \right\rangle $, given in (\ref{eq2}), spans a three
dimensional subspace (with an orthonormal basis $\left\{  |00\rangle
,|11\rangle,\left\vert \chi\left(  \theta,{\theta}_{0}\right)  \right\rangle
\right\}  $) of the total four dimensional two-qubit Hilbert space.

For estimation, let us choose a POVM $\mathcal{M}=\left\{  E_{r}:r\in
\Lambda\right\}  $ on the linear span $\mathcal{L}\left(  S_{1,1}\left(
\theta,{\theta}_{0}\right)  \right)  $ of $S_{1,1}\left(  \theta,{\theta}%
_{0}\right)  $, where
\begin{equation}
E_{r}=C_{r}P\left[  {\alpha}_{1r}|00\rangle+{\alpha}_{2r}|11\rangle+{\alpha
}_{3r}\left\vert \chi\left(  \theta,{\theta}_{0}\right)  \right\rangle
\right]  , \label{eq4}%
\end{equation}
with
\begin{equation}%
\begin{array}
[c]{lcl}%
C_{r} & > & 0,\\
\sum_{j=1}^{3}\left\vert {\alpha}_{jr}\right\vert ^{2} & = & 1~~\mathrm{for}%
~\mathrm{all}~r\in\Lambda,\\
\sum_{r\in\Lambda}C_{r}{\alpha}_{jr}{\alpha}_{kr}^{\ast} & = & {\delta}%
_{jk}~~\mathrm{for}~\mathrm{all}~j,k=1,2,3.
\end{array}
\label{eq5}%
\end{equation}
The score is
\[
s(\mathcal{M},r,(\theta,\phi))=\left\vert \left\langle \psi\left(  \theta
,\phi\right)  |\psi\left(  \theta,{\phi}_{r}\right)  \right\rangle \right\vert
^{2}.
\]
So, the average state estimation fidelity is given by
\[
{\overline{F}}_{1,1}\left(  \theta,{\theta}_{0}\right)  =\frac{1}{2\pi}%
\sum_{r\in\Lambda}\int_{\phi=0}^{2\pi}\left\langle \Psi\left(  \theta,{\theta
}_{0},\phi\right)  \right\vert E_{r}\left\vert \Psi\left(  \theta,{\theta}%
_{0},\phi\right)  \right\rangle \left\vert \left\langle \psi\left(
\theta,\phi\right)  |\psi\left(  \theta,{\phi}_{r}\right)  \right\rangle
\right\vert ^{2}d\phi
\]%
\[
=1-\frac{\mathrm{sin}^{2}\theta}{2}+\frac{\mathrm{sin}^{2}\theta}{2}N\left(
\theta,{\theta}_{0}\right)  \times
\]%
\begin{equation}
\left\{  \mathrm{cos}\frac{\theta}{2}~\mathrm{cos}\frac{\theta+{\theta}_{0}%
}{2}\sum_{r\in\Lambda}C_{r}~\mathrm{Re}\left\{  {\alpha}_{1r}{\alpha}%
_{3r}^{\ast}e^{i{\phi}_{r}}\right\}  +~\mathrm{sin}\frac{\theta}%
{2}~\mathrm{sin}\frac{\theta+{\theta}_{0}}{2}\sum_{r\in\Lambda}C_{r}%
~\mathrm{Re}\left\{  {\alpha}_{2r}{\alpha}_{3r}^{\ast}e^{-i{\phi}_{r}%
}\right\}  \right\}  .
\end{equation}
Thus we see that
\begin{equation}
{\overline{F}}_{1,1}\left(  \theta,{\theta}_{0}\right)  \leq1-\frac
{\mathrm{sin}^{2}\theta}{2}+\frac{\mathrm{sin}^{2}\theta~\mathrm{cos}%
\frac{{\theta}_{0}}{2}}{2}N\left(  \theta,{\theta}_{0}\right)  , \label{eq6}%
\end{equation}
a quantity independent of the choice of the POVM. Note that as here
$-\frac{\pi}{2}\leq-\frac{\theta}{2}\leq\frac{{\theta}_{0}}{2}\leq\frac
{\pi-\theta}{2}\leq\frac{\pi}{2}$, therefore $\mathrm{cos}\frac{{\theta}_{0}%
}{2}\geq0$.

The following is a choice for which equality holds good in (\ref{eq6}):
\[
\Lambda=\{1,2,3\},~~C_{r}=1~~\mathrm{for}~\mathrm{all}~r\in\Lambda,
\]%
\begin{equation}
{\alpha}_{jr}=\frac{1}{\sqrt{3}}~\mathrm{exp}\left[  \frac{2\pi i(j-1)(r-1)}%
{3}\right]  ~~\mathrm{for}~\mathrm{all}~r\in\Lambda~~\mathrm{and}%
~\mathrm{for}~\mathrm{all}~j=1,2,3, \label{eq7}%
\end{equation}
while
\begin{equation}
{\phi}_{1}=0,~{\phi}_{2}=\frac{4\pi}{3},~{\phi}_{3}=\frac{2\pi}{3}.
\label{eq7.1}%
\end{equation}
Thus we see that the optimal average fidelity, in this case, is given by
\begin{equation}
{\overline{F}}_{1,1}^{\mathrm{max}}\left(  \theta,{\theta}_{0}\right)
=1-\frac{\mathrm{sin}^{2}\theta}{2}+\frac{\mathrm{sin}^{2}\theta
~\mathrm{cos}\frac{{\theta}_{0}}{2}}{2}N\left(  \theta,{\theta}_{0}\right)  ,
\label{eq8}%
\end{equation}
where $N\left(  \theta,{\theta}_{0}\right)  $ is given in (\ref{eq3}). We
would like to know now for which value(s) of ${\theta}_{0}\in\lbrack
-\theta,\pi-\theta]$, ${\overline{F}}_{1,1}^{\mathrm{max}}\left(
\theta,{\theta}_{0}\right)  $ is maximum, given any arbitrary but fixed
$\theta\in\lbrack0,\pi]$. Note that (according to our notations, used in
earlier sections)
\[%
\begin{tabular}
[c]{lll}%
${\overline{F}}_{2,0}^{\mathrm{max}}(\theta)={\overline{F}}_{1,1}%
^{\mathrm{max}}(\theta,0)$ & and & ${\overline{F}}_{1,1}^{\mathrm{max}}%
(\theta)={\overline{F}}_{1,1}^{\mathrm{max}}(\theta,\pi-2\theta).$%
\end{tabular}
\]
Also note that the maximum value of ${\overline{F}}_{1,1}^{\mathrm{max}%
}\left(  \theta,{\theta}_{0}\right)  $ is equal to $1$ for both $\theta=0$ as
well as $\theta=\pi$ (irrespective of values of ${\theta}_{0}$). Basically,
when $\theta=0$, the state estimation (which we considered here) reduces to
estimating the direction of the qubit $|0\rangle$, given the supply of the
two-qubit states $|0\rangle\otimes\left\vert \psi\left(  {\theta}_{0}%
,\phi\right)  \right\rangle $ for the uniform distribution of $\phi$ over
$[0,2\pi)$. Hence the optimal fidelity must be 1. Same is the case when
$\theta=\pi$. On the other hand, for given any $\theta\in\lbrack0,\pi]$,
${\overline{F}}_{1,1}^{\mathrm{max}}\left(  \theta,{\theta}_{0}\right)  $ will
reach its minimum when ${\theta}_{0}=-\theta$ and ${\theta}_{0}=\pi-\theta$.
This is because when ${\theta}_{0}=-\theta$ (or ${\theta}_{0}=\pi-\theta$),
the set of supplied two-qubit states is of the form $\{|\psi(\theta
,\phi)\rangle\otimes|0\rangle:\phi\in\lbrack0,2\pi)\}$ (or $\{|\psi
(\theta,\phi)\rangle\otimes|1\rangle:\phi\in\lbrack0,2\pi)\}$). And this gives
rise to the same optimal average fidelity ${\overline{F}}_{1,1}^{\mathrm{max}%
}\left(  \theta,-\theta\right)  $ ($={\overline{F}}_{1,1}^{\mathrm{max}%
}\left(  \theta,\pi-{\theta}\right)  $) as in the case of estimating the
direction of the qubits $|\psi(\theta,\phi)\rangle$ (for fixed $\theta$), when
the supplied set of states is the circle $S_{\theta}=\{|\psi(\theta
,\phi)\rangle:\phi\in\lbrack0,2\pi)\}$ itself. So, in our notation, we have
\begin{equation}
{\overline{F}}_{1,0}^{\mathrm{max}}(\theta)={\overline{F}}_{1,1}%
^{\mathrm{max}}\left(  \theta,{\theta}_{0}=-{\theta}\right)  ={\overline{F}%
}_{1,1}^{\mathrm{max}}\left(  \theta,{\theta})_{0}=\pi-{\theta}\right)
=\mathrm{min}\left\{  {\overline{F}}_{1,1}^{\mathrm{max}}\left(
\theta,{\theta}_{0}\right)  :{\theta}_{0}\in\lbrack-\theta,\pi-\theta
]\right\}  \label{eq8'}%
\end{equation}
Again
\[
{\overline{F}}_{1,1}^{\mathrm{max}}\left(  \theta=\frac{\pi}{2},{\theta}%
_{0}\right)  =\frac{1}{2}+\frac{\mathrm{cos}\frac{{\theta}_{0}}{2}}{2\sqrt{2}%
},
\]
which will take its maximum value $\frac{1}{2}+\frac{1}{2\sqrt{2}}$ for
${\theta}_{0}=0$. Thus for estimating the direction of pure qubit, uniformly
distributed on a given great circle, if two pure qubits are supplied, it is
always better to supply two parallel qubits (or, equivalently two
anti-parallel qubits), rather than supplying one pure qubit from the great
circle and another corresponding qubit from a small circle whose plane is
parallel to that of the great circle.

In the maximization procedure of ${\overline{F}}_{1,1}^{\mathrm{max}}\left(
\theta,{\theta}_{0}\right)  $ over all values of ${\theta}_{0}\in
\lbrack-\theta,\pi-\theta]$, we therefore can assume that $\theta$ is
different from $0$, $\frac{\pi}{2}$, and $\pi$. In Figure 7, $\overline
{F}_{1,1}^{\max}(\theta,\theta_{0})$ is plotted for $0\leq\theta\leq\pi$ and
$-\theta\leq\theta_{0}\leq\pi-\theta$.
\begin{center}
\includegraphics[
height=2.3565in,
width=3.1712in
]%
{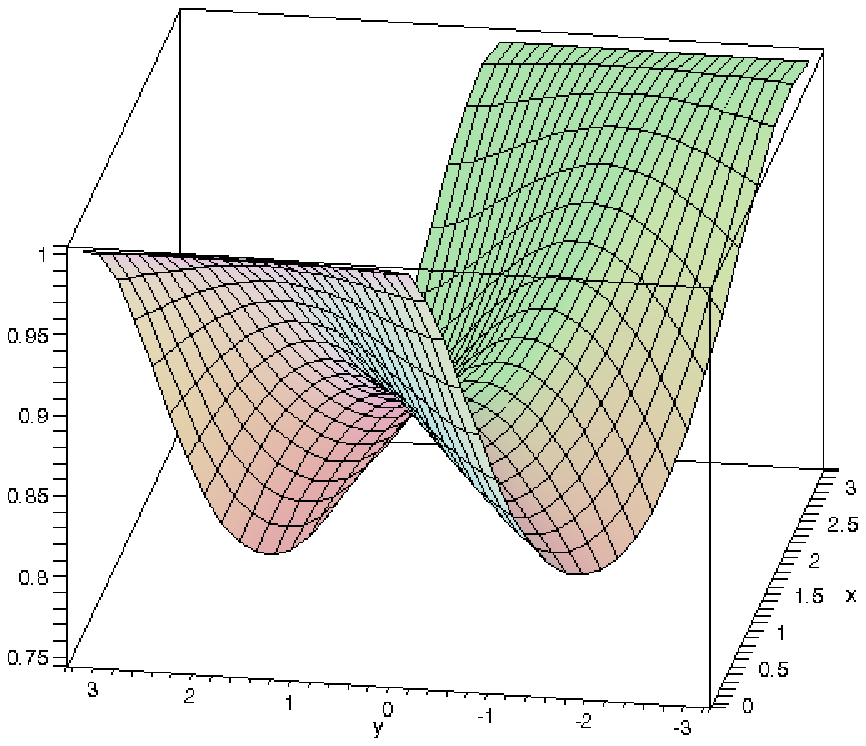}%
\\
Figure 7: The plot of $\overline{F}_{1,1}^{\max}(\theta,\theta_{0})$ for
$0\leq\theta\leq\pi$ and $-\theta\leq\theta_{0}\leq\pi-\theta$. The points of
$\overline{F}_{1,1}^{\max}(\theta,0)$ are on the intersection of $\overline
{F}_{1,1}^{\max}(\theta,\theta_{0})$ with the plane $y=0$. The points of
$\overline{F}_{1,1}^{\max}(\theta,\pi-2\theta)$ are on the intersection of
$\overline{F}_{1,1}^{\max}(\theta,\theta_{0})$ with the plane $2x+y=\pi$.
\label{3d}%
\end{center}

Our idea behind the choice of the set $S_{1,1}\left(  \theta,{\theta}%
_{0}\right)  $, from which a two-qubit state has to be supplied for the
estimation, is to check whether for any fixed $\theta\in\left(  \lbrack
0,\pi]-\left\{  0,\frac{\pi}{2},\pi\right\}  \right)  $, states from
$S_{1,1}\left(  \theta,0\right)  $ give the minimum value of ${\overline{F}%
}_{1,1}^{\mathrm{max}}\left(  \theta,{\theta}_{0}\right)  $ and states from
$S_{1,1}\left(  \theta,\pi-2\theta\right)  $ give the maximum value of
${\overline{F}}_{1,1}^{\mathrm{max}}\left(  \theta,{\theta}_{0}\right)  $. But
one can check that for all $\theta\in\lbrack0,\pi]$,
\begin{equation}
\left[  \frac{\partial{{\overline{F}}_{1,1}^{\mathrm{max}}\left(
\theta,{\theta}_{0}\right)  }}{\partial{{\theta}_{0}}}\right]  _{{\theta}%
_{0}=0}=\frac{\mathrm{sin}2\theta}{4}, \label{eq9}%
\end{equation}
and
\begin{equation}
\left[  \frac{\partial{{\overline{F}}_{1,1}^{\mathrm{max}}\left(
\theta,{\theta}_{0}\right)  }}{\partial{{\theta}_{0}}}\right]  _{{\theta}%
_{0}=\pi-2\theta}=-\frac{1}{2}~\mathrm{cos}^{2}\theta~\mathrm{sin}2\theta.
\label{eq10}%
\end{equation}
The right hand sides of both (\ref{eq9}) and (\ref{eq10}) are equal to zero if
and only if $\theta=0,\frac{\pi}{2},\pi$. Thus for given any $\theta\in\left(
\lbrack0,\pi]-\left\{  0,\frac{\pi}{2},\pi\right\}  \right)  $, neither the
minimum value of ${\overline{F}}_{1,1}^{\mathrm{max}}\left(  \theta,{\theta
}_{0}\right)  $ is attained by the supply of parallel qubits $\left\vert
{\Psi}_{2,0}(\theta,\phi)\right\rangle $, nor the maximum value of
${\overline{F}}_{1,1}^{\mathrm{max}}\left(  \theta,{\theta}_{0}\right)  $ is
attained by the supply of anti-parallel qubits $\left\vert {\Psi}_{1,1}%
(\theta,\phi)\right\rangle $. This shows that in order to extract best
information about the direction of the Bloch vector $\widehat{\mathbf{n}%
}=(\mathrm{\sin}\theta\cos\phi,\sin\theta\sin\phi,\sin\theta)$ of a qubit
$|\psi(\theta,\phi)\rangle=\cos\frac{\theta}{2}|0\rangle+e^{i\phi}\sin
\frac{\theta}{2}|1\rangle$ (contained within the circle $S_{\theta}%
=\{|\psi(\theta,\phi)\rangle:\phi\in\lbrack0,2\pi)\}$), we need to encode the
direction of the Bloch vector in a two-qubit pure state (\emph{i.e.} the
supplied state) in a form which in general is neither parallel nor anti-parallel.

\section{Estimation of qubits from two diametrically opposite circles}

We have seen that in the case of estimating the direction of the Bloch vector
$\widehat{\mathbf{n}}=(\mathrm{\sin}\theta\cos\phi,\sin\theta\sin\phi
,\cos\theta)$ of the qubit $|\psi(\theta,\phi)\rangle\in S_{\theta}%
=\{|\psi(\theta,\phi)\rangle:\phi\in\lbrack0,2\pi)\}$, the anti-parallel
qubits $\left\vert {\Psi}_{1,1}\right\rangle =|\psi(\theta,\phi)\rangle
\otimes|\psi(\pi-\theta,\pi+\phi)\rangle$ give better information compared to
the parallel qubits $\left\vert {\Psi}_{2,0}\right\rangle =|\psi(\theta
,\phi)\rangle\otimes|\psi(\theta,\phi)\rangle$ (where $\phi$ is uniformly
distributed over $[0.2\pi)$), even though both anti-parallel as well as
parallel qubits, in this scenario, span three dimensional subspaces. By
symmetry, it can be shown that in the case of estimating the direction of the
Bloch vector
\[
\widehat{\mathbf{m}}\left(  \widehat{\mathbf{n}}\right)  =(\mathrm{\sin}%
(\pi-\theta)\cos\phi,\sin(\pi-\theta)\sin\phi,\cos(\pi-\theta))=(\mathrm{\sin
}\theta\cos\phi,\sin\theta\sin\phi,-\mathrm{\cos}\theta),
\]
the anti-parallel qubits $\left\vert {\Psi}_{1,1}\right\rangle =|\psi
(\theta,\phi)\rangle\otimes|\psi(\pi-\theta,\pi+\phi)\rangle$ give better
information compared to the parallel qubits $\left\vert {\Psi}_{2,0}%
\right\rangle =|\psi(\theta,\phi)\rangle\otimes|\psi(\theta,\phi)\rangle$
(where $\phi$ is uniformly distributed over $[0.2\pi)$) (see Figure 8). It
should be noted here that the score of the game in the former case is
$\frac{1+\widehat{\mathbf{n}}\cdot\widehat{\mathbf{n}}_{r}}{2}$, while, for
the later case, it is equal to $\frac{1+\widehat{{\mathbf{m}}}\left(
\widehat{\mathbf{n}}\right)  \cdot\widehat{{\mathbf{m}}}\left(  \widehat
{\mathbf{n}}_{r}\right)  }{2}$, where $\widehat{\mathbf{n}}_{r}=(\mathrm{\sin
}\theta\cos{\phi}_{r},\sin\theta\sin{\phi}_{r},\cos\theta)$. Hence,
$\frac{1+\widehat{\mathbf{n}}\cdot\widehat{\mathbf{n}}_{r}}{2}=\frac
{1+\widehat{{\mathbf{m}}}\left(  \widehat{\mathbf{n}}\right)  \cdot
\widehat{{\mathbf{m}}}\left(  \widehat{\mathbf{n}}_{r}\right)  }{2}$. Consider
now the problem of estimating the direction of the Bloch vector $\widehat
{\mathbf{n}}=(\mathrm{\sin}\theta\cos\phi,\sin\theta\sin\phi,\cos\theta)$
associated to the qubit $|\psi(\theta,\phi)\rangle=\cos\frac{\theta}%
{2}|0\rangle+e^{i\phi}\sin\frac{\theta}{2}|1\rangle\in\{|\psi(\theta
,\phi)\rangle:\phi\in\lbrack0,2\pi)\}$, when the supplied two qubits can be
either of the form $\left\vert {\Psi}_{2,0}(\theta,\phi)\right\rangle
=|\psi(\theta,\phi)\rangle\otimes|\psi(\theta,\phi)\rangle$ or of the form
$\left\vert {\Psi}_{2,0}(\pi-\theta,\phi)\right\rangle =|\psi(\pi-\theta
,\phi)\rangle\otimes|\psi(\pi-\theta,\phi)\rangle$, in the case of parallel
qubits, while the supplied two qubits can be either of the form $\left\vert
{\Psi}_{1,1}(\theta,\phi)\right\rangle =|\psi(\theta,\phi)\rangle\otimes
|\psi(\pi-\theta,\pi+\phi)\rangle$ or of the form $\left\vert {\Psi}_{1,1}%
(\pi-\theta,\phi)\right\rangle =|\psi(\pi-\theta,\phi)\rangle\otimes
|\psi(\theta,\pi+\phi)\rangle$, in the case of anti-parallel qubits.%

\begin{center}
\includegraphics[
height=2.1257in,
width=2.3173in
]%
{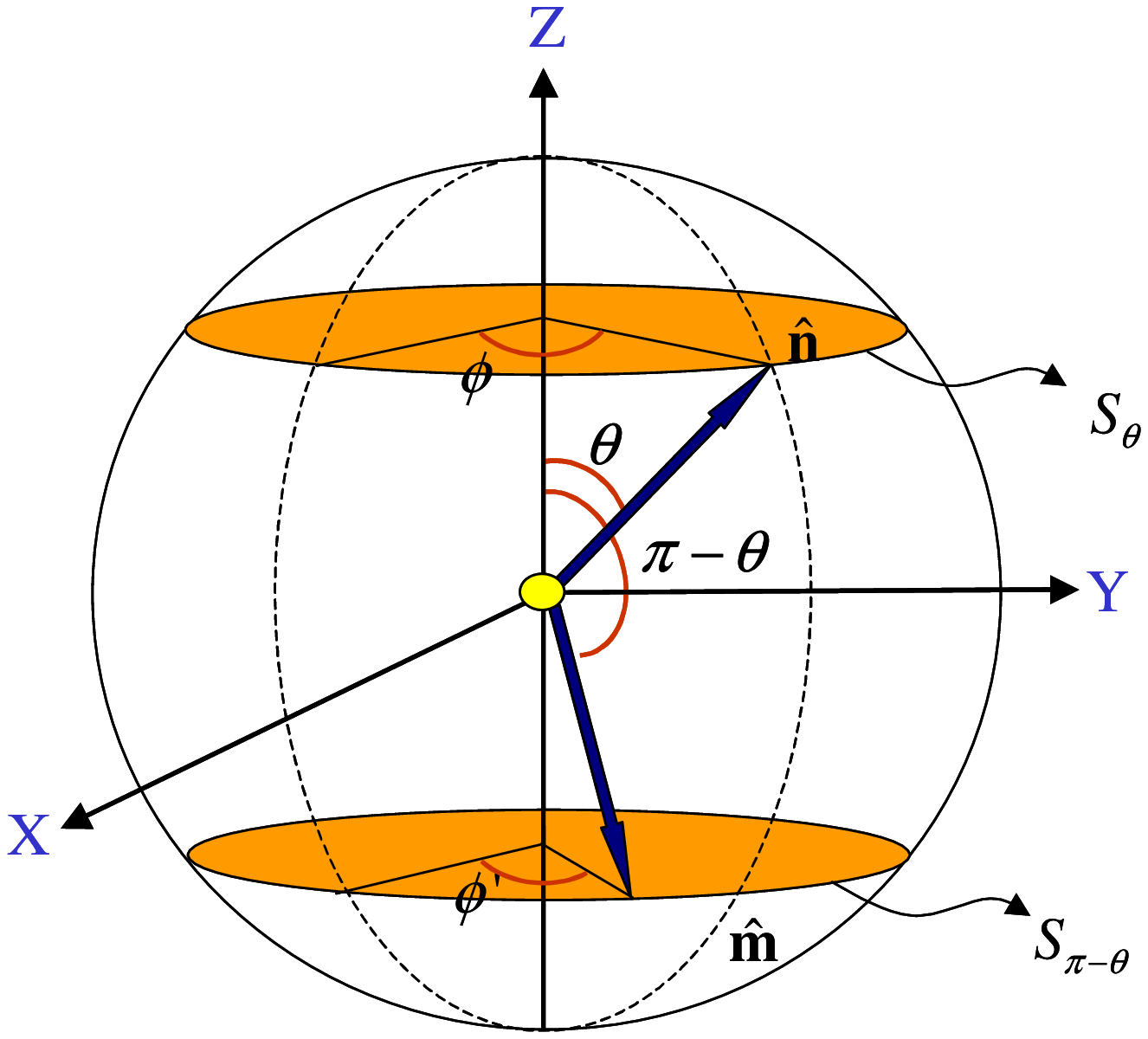}%
\\
Figure 8:
\label{2vectors}%
\end{center}
Note that the dimension of the linear spans of the sets of all parallel and
anti-parallel qubits, $|\psi\rangle\otimes|\psi\rangle$ and $|\psi
\rangle\otimes|{\psi}^{\bot}\rangle$, are respectively \emph{three} and
\emph{four}, whenever the qubit $|\psi\rangle$ is taken from the set of two
diametrically opposite circles
\begin{equation}
S_{\theta,\pi-\theta}=\{|\psi(\theta,\phi)\rangle:\phi\in\lbrack0,2\pi
)\}\cup\{|\psi(\pi-\theta,\phi)\rangle:\phi\in\lbrack0,2\pi)\}, \label{twocir}%
\end{equation}
where $\theta\in\lbrack0,\pi]-\left\{  0,\frac{\pi}{2},\pi\right\}  $ is
arbitrary but fixed. (The bases of the spans are in fact $\{|00\rangle
,|11\rangle,\frac{1}{\sqrt{2}}(|01\rangle+|10\rangle)\}$ and $\{|00\rangle
,|11\rangle,|01\rangle,|10\rangle+|10\rangle\}$, respectively.) So, according
to dimensional argument, anti-parallel qubits should give more information
about the direction of the Bloch vector of the qubit $|\psi\rangle$, compared
to parallel qubits. The question is here to choose an appropriate score. The
general estimation strategy is then as follows. For any qubit $|\psi
\rangle=\left\vert \psi\left(  {\theta}^{\prime},{\phi}^{\prime}\right)
\right\rangle $ (where ${\theta}^{\prime}\in\lbrack0,\pi]$ and ${\phi}%
^{\prime}\in\lbrack0,2\pi)$), we denote by $\left\vert {\psi}^{\bot
}\right\rangle $ the corresponding orthogonal qubit $\left\vert \psi\left(
\pi-{\theta}^{\prime},\pi+{\phi}^{\prime}\right)  \right\rangle $. Let%
\begin{equation}%
\begin{tabular}
[c]{lll}%
$S_{\theta,\pi-\theta}^{||}=\left\{  |\psi\rangle\otimes|\psi\rangle
:|\psi\rangle\in S_{\theta,\pi-\theta}\right\}  $ & and & $S_{\theta
,\pi-\theta}^{\bot}=\left\{  |\psi\rangle\otimes|{\psi}^{\bot}\rangle
:|\psi\rangle\in S_{\theta,\pi-\theta}\right\}  ,$%
\end{tabular}
\label{paraanti}%
\end{equation}
where $S_{\theta,\pi-\theta}$ is given in \ref{twocir}. If a state $|{\Psi
}^{||}\rangle\equiv|\psi\rangle\otimes|\psi\rangle$ is supplied from
$S_{\theta,\pi-\theta}^{||}$, we perform a POVM $\mathcal{M}=\left\{
A_{r}=C_{r}P\left[  {\alpha}_{1r}|00\rangle+{\alpha}_{2r}|11\rangle+{\alpha
}_{3r}|{\psi}^{+}\rangle\right]  :r\in\Lambda\right\}  $ on this state (where
$|{\psi}^{+}\rangle=\frac{1}{\sqrt{2}}(|01\rangle+|10\rangle)$); if the $r$-th
outcome of the measurement occurs (with probability $\langle{\Psi}^{||}%
|A_{r}|{\Psi}^{||}\rangle$), the estimated qubit is taken as the density
matrix ${\rho}_{r}$ (and hence, the score is $s(\mathcal{M},\mathcal{T}%
=\{{\rho}_{r}:r\in\Lambda\},|\psi\rangle)=\left\langle \psi\right\vert {\rho
}_{r}\left\vert \psi\right\rangle $). On the other hand, if a state $|{\Psi
}^{\bot}\rangle\equiv|\psi\rangle\otimes|{\psi}^{\bot}\rangle$ is supplied
from $S_{\theta,\pi-\theta}^{\bot}$, we perform a POVM $\mathcal{M}=\left\{
A_{r}=C_{r}P\left[  {\alpha}_{1r}|00\rangle+{\alpha}_{2r}|11\rangle+{\alpha
}_{3r}|01\rangle+{\alpha}_{4r}|10\rangle\right]  :r\in\Lambda\right\}  $ on
this state (where $\left\vert {\psi}^{+}\right\rangle =\frac{1}{\sqrt{2}%
}(|01\rangle+|10\rangle)$); and if the $r$-th outcome of the measurement
occurs (with probability $\langle{\Psi}^{\bot}|A_{r}|{\Psi}^{\bot}\rangle$),
the estimated qubit is taken as the density matrix ${\rho}_{r}$ (and hence,
the score is $s(\mathcal{M},\mathcal{T}=\{{\rho}_{r}:r\in\Lambda
\},|\psi\rangle)=\left\langle \psi\right\vert {\rho}_{r}\left\vert
\psi\right\rangle $). Thus the average fidelity of estimation for parallel and
anti-parallel qubits are respectively given by
\[
{\overline{F}}_{\theta,\pi-\theta;||}(\mathcal{M},\mathcal{T})=\frac{1}%
{2}\times\frac{1}{2\pi}\int_{\phi=0}^{2\pi}\left\{  \sum_{r\in\Lambda
}\left\langle {\Psi}_{2,0}(\theta,\phi)\right\vert A_{r}\left\vert {\Psi
}_{2,0}(\theta,\phi)\right\rangle \left\langle \psi(\theta,\phi)\right\vert
{\rho}_{r}\left\vert \psi(\theta,\phi)\right\rangle \right\}  d\phi+
\]%
\begin{equation}
\frac{1}{2}\times\frac{1}{2\pi}\int_{\phi=0}^{2\pi}\left\{  \sum_{r\in\Lambda
}\left\langle {\Psi}_{2,0}(\pi-\theta,{\phi}^{\prime})\right\vert
A_{r}\left\vert {\Psi}_{2,0}(\pi-\theta,{\phi}^{\prime})\right\rangle
\left\langle \psi(\pi-\theta,{\phi}^{\prime})\right\vert {\rho}_{r}\left\vert
\psi(\pi-\theta,{\phi}^{\prime})\right\rangle \right\}  d{\phi}^{\prime},
\label{twocirfip}%
\end{equation}
and
\[
{\overline{F}}_{\theta,\pi-\theta;\bot}(\mathcal{M},\mathcal{T})=\frac{1}%
{2}\times\frac{1}{2\pi}\int_{\phi=0}^{2\pi}\left\{  \sum_{r\in\Lambda
}\left\langle {\Psi}_{1,1}(\theta,\phi)\right\vert A_{r}\left\vert {\Psi
}_{1,1}(\theta,\phi)\right\rangle \left\langle \psi(\theta,\phi)\right\vert
{\rho}_{r}\left\vert \psi(\theta,\phi)\right\rangle \right\}  d\phi+
\]%
\begin{equation}
\frac{1}{2}\times\frac{1}{2\pi}\int_{\phi=0}^{2\pi}\left\{  \sum_{r\in\Lambda
}\left\langle {\Psi}_{1,1}(\pi-\theta,{\phi}^{\prime})\right\vert
A_{r}\left\vert {\Psi}_{1,1}(\pi-\theta,{\phi}^{\prime})\right\rangle
\left\langle \psi(\pi-\theta,{\phi}^{\prime})\right\vert {\rho}_{r}\left\vert
\psi(\pi-\theta,{\phi}^{\prime})\right\rangle \right\}  d{\phi}^{\prime}.
\label{twocirfia}%
\end{equation}
Since our motivation is to estimate the direction of the Bloch vector of the
qubit taken from $S_{\theta,\pi-\theta}$, the estimated qubit ${\rho}_{r} $
should be of the form
\begin{equation}
{\rho}_{r}={\lambda}_{r}\left\vert \psi\left(  \theta,{\phi}_{r}\right)
\right\rangle \left\langle \psi\left(  \theta,{\phi}_{r}\right)  \right\vert
+\left(  1-{\lambda}_{r}\right)  \left\vert \psi\left(  \pi-\theta,{\phi}%
_{r}^{\prime}\right)  \right\rangle \left\langle \psi\left(  \pi-\theta,{\phi
}_{r}^{\prime}\right)  \right\vert , \label{estimatedstate}%
\end{equation}
where $0\leq{\lambda}_{r}\leq1$, ${\phi}_{r},{\phi}_{r}^{\prime}\in
\lbrack0,2\pi)$. The parameters ${\lambda}_{r}$, ${\phi}_{r}$, ${\phi}%
_{r}^{\prime}$ need to be chosen in such a way that average state estimation
fidelities would become maximum for the given POVM $\mathcal{M}$. For our
purpose, we take ${\lambda}_{r}=1$. The reason behind this choice is the
following: the optimal state estimation fidelity for estimating the direction
of the Bloch vector $\widehat{\mathbf{n}}=(\mathrm{sin}\theta~\mathrm{cos}%
\phi,~\mathrm{sin}\theta~\mathrm{sin}\phi,~\mathrm{cos}\theta)$ of the qubit
$|\psi(\theta,\phi)\rangle\in S_{\theta}=\{|\psi(\theta,\phi)\rangle:\phi
\in\lbrack0,2\pi)\}$, when the supplied state is $\left\vert {\Psi}%
_{2,0}(\theta,\phi)\right\rangle $, is the same as for estimating the
direction of the Bloch vector $\widehat{\mathbf{m}}\left(  \widehat
{\mathbf{n}}\right)  =(\mathrm{sin}\theta~\mathrm{cos}\phi,~\mathrm{sin}%
\theta~\mathrm{sin}\phi,~-\mathrm{cos}\theta)$, even when the supplied state
is as above. This is also true for anti-parallel states. It follows that, the
maximum values of
\[
\frac{1}{2\pi}\int_{\phi=0}^{2\pi}\sum_{r\in\Lambda}\left\langle {\Psi}%
_{j,k}(\theta,\phi)\right\vert A_{r}\left\vert {\Psi}_{j,k}(\theta
,\phi)\right\rangle \left\vert \left\langle \psi(\theta,\phi)|\psi\left(
\theta,{\phi}_{r}\right)  \right\rangle \right\vert ^{2}d\phi
\]
and
\[
\frac{1}{2\pi}\int_{\phi=0}^{2\pi}\sum_{r\in\Lambda}\left\langle {\Psi}%
_{j,k}(\theta,\phi)\right\vert A_{r}\left\vert {\Psi}_{j,k}(\theta
,\phi)\right\rangle \left\vert \left\langle \psi(\pi-\theta,{\phi}^{\prime
})|\psi\left(  \pi-\theta,{\phi}_{r}\right)  \right\rangle \right\vert
^{2}d{\phi}^{\prime},
\]
where $(j,k)\in\{(2,0),(1,1)\}$, are equal.

\subsection{Parallel case}

With the choice of the estimated state ${\rho}_{r}=P[\left\vert \psi\left(
\theta,{\phi}_{r}\right)  \right\rangle ]$, for the $r$-th measurement outcome
of the POVM
\[
\mathcal{M}=\left\{  E_{r}=C_{r}P\left[  {\alpha}_{1r}|00\rangle+{\alpha}%
_{2r}|11\rangle+{\alpha}_{3r}\left\vert {\psi}^{+}\right\rangle \right]
:r\in\Lambda\right\}  ,
\]
the average fidelity when parallel qubits are supplied is%
\begin{align}
{\overline{F}}_{\theta,\pi-\theta;||}(\mathcal{M},\mathcal{T})  &  =\frac
{1}{2}\left(  1+\frac{\left(  2-\cos\theta\right)  \sin^{3}\theta}{4\sqrt{2}%
}\sum_{r\in\Lambda}C_{r}|\alpha_{1r}\alpha_{3r}|\cos\left(  \varepsilon
_{1r}-\varepsilon_{3r}+{\phi}_{r}\right)  \right. \label{pafi}\\
&  +\left.  \frac{\left(  2+\cos\theta\right)  \sin^{3}\theta}{4\sqrt{2}}%
\sum_{r\in\Lambda}C_{r}|\alpha_{2r}\alpha_{3r}|\cos\left(  \varepsilon
_{2r}-\varepsilon_{3r}-{\phi}_{r}\right)  \right)  ,\nonumber
\end{align}
where $\alpha_{jr}=|\alpha_{jr}|e^{i\varepsilon_{jr}}$ for $j=1,2,3$ and
$r\in\Lambda$. We have to maximize ${\overline{F}}_{\theta,\pi-\theta
;||}(\mathcal{M},\mathcal{T})$ over all possible choices of $\mathcal{M}$ and
$\mathcal{T}$, and subject to the constraints A, B, and C, that is:
\begin{equation}%
\begin{tabular}
[c]{l}%
$C_{r}>0\text{ for all }r\in\Lambda,$\\
$\sum_{j=1}^{3}|\alpha_{jr}|^{2}=1\text{ for all }r\in\Lambda$\\
$\sum_{r\in\Lambda}C_{r}\alpha_{jr}\alpha_{kr}^{\ast}=\delta_{jk}\text{ for
all }j,k=1,2,3.$%
\end{tabular}
\label{parab}%
\end{equation}
From (\ref{pafi}) it follows that
\begin{align}
{\overline{F}}_{\theta,\pi-\theta;||}(\mathcal{M},\mathcal{T})  &  \leq
\frac{1}{2}\left[  1+\frac{\left(  2-\cos\theta\right)  \sin^{3}\theta}%
{4\sqrt{2}}\left(  \sum_{r\in\Lambda}C_{r}|\alpha_{1r}|^{2}\right)
^{1/2}\right.  \left(  \sum_{r\in\Lambda}C_{r}|\alpha_{3r}|^{2}\right)
^{1/2}\nonumber\\
&  +\left.  \frac{\left(  2+\cos\theta\right)  \sin^{3}\theta}{4\sqrt{2}%
}\left(  \sum_{r\in\Lambda}C_{r}|\alpha_{2r}|^{2}\right)  ^{1/2}\left(
\sum_{r\in\Lambda}C_{r}|\alpha_{3r}|^{2}\right)  ^{1/2}\right] \nonumber\\
&  =\frac{1}{2}\left(  1+\frac{\left(  2-\cos\theta\right)  \sin^{3}\theta
}{4\sqrt{2}}+\frac{\left(  2+\cos\theta\right)  \sin^{3}\theta}{4\sqrt{2}%
}\right) \nonumber\\
&  =\frac{1}{2}\left(  1+\frac{\sin^{3}\theta}{\sqrt{2}}\right)  .
\label{optpara}%
\end{align}
Let us chose
\begin{equation}%
\begin{tabular}
[c]{l}%
$\Lambda=\{1,2,3\},$\\
$C_{r}=1\text{ for all }r\in\Lambda,$\\
$\alpha_{11}=\alpha_{12}=\alpha_{21}=\alpha_{31}=\alpha_{13}=\frac{1}{\sqrt
{3}},\alpha_{22}=\alpha_{33}=\frac{e^{4\pi i/3}}{\sqrt{3}},\alpha_{32}%
=\alpha_{23}=\frac{e^{2\pi i/3}}{\sqrt{3}},$\\
$\phi_{1}=0,\phi_{2}=\frac{2\pi}{3},\phi_{3}=\frac{4\pi}{3}.$%
\end{tabular}
\label{parachoi}%
\end{equation}
For the choice (\ref{parachoi}), one can see that all the conditions in
(\ref{parab}) are satisfied, then%
\begin{equation}
{\overline{F}}_{\theta,\pi-\theta;||}(\mathcal{M},\mathcal{T})=\frac{1}%
{2}\left(  1+\frac{\sin^{3}\theta}{\sqrt{2}}\right)  ={\overline{F}}%
_{\theta,\pi-\theta;||}^{\max}. \label{paramax}%
\end{equation}
The elements of the POVM can be expressed in the following matrix in terms of
the basis $\{|00\rangle,|11\rangle,|{\psi}^{+}\rangle\}$:%
\[
\left[
\begin{array}
[c]{ccc}%
{\alpha}_{11}=\frac{1}{\sqrt{3}} & {\alpha}_{21}=\frac{1}{\sqrt{3}} & {\alpha
}_{31}=\frac{1}{\sqrt{3}}\\
{\alpha}_{12}=\frac{1}{\sqrt{3}} & {\alpha}_{22}=\frac{e^{4\pi i/3}}{\sqrt{3}}
& {\alpha}_{32}=\frac{e^{2\pi i/3}}{\sqrt{3}}\\
{\alpha}_{13}=\frac{1}{\sqrt{3}} & {\alpha}_{23}=\frac{e^{2\pi i/3}}{\sqrt{3}}
& {\alpha}_{33}=\frac{e^{4\pi i/3}}{\sqrt{3}}%
\end{array}
\right]  .
\]
This matrix is the discrete Fourier transform of dimension $3$.

\subsection{Anti-parallel case}

With the choice of the estimated state ${\rho}_{r}=P[\left\vert \psi\left(
\theta,{\phi}_{r}\right)  \right\rangle ]$, for the $r$-th measurement outcome
of the POVM
\[
\mathcal{M}=\left\{  E_{r}=C_{r}P\left[  {\alpha}_{1r}|00\rangle+{\alpha}%
_{2r}|11\rangle+{\alpha}_{3r}\left\vert 01\right\rangle +{\alpha}%
_{4r}\left\vert 10\right\rangle \right]  :r\in\Lambda\right\}  ,
\]
the average fidelity when anti-parallel qubits are supplied is%
\begin{align}
{\overline{F}}_{\theta,\pi-\theta;\bot}(\mathcal{M},\mathcal{T})  &  =\frac
{1}{2}\left[  1-\frac{\sin^{3}\theta}{4}\sum_{r\in\Lambda}C_{r}|\alpha
_{1r}\alpha_{3r}|\cos\left(  \varepsilon_{1r}-\varepsilon_{3r}+{\phi}%
_{r}\right)  \right. \label{bantifi}\\
&  +\frac{\sin^{3}\theta}{4}\sum_{r\in\Lambda}C_{r}|\alpha_{1r}\alpha
_{4r}|\cos\left(  \varepsilon_{1r}-\varepsilon_{4r}+{\phi}_{r}\right)
\nonumber\\
&  +\frac{\sin^{3}\theta}{4}\sum_{r\in\Lambda}C_{r}|\alpha_{2r}\alpha
_{3r}|\cos\left(  \varepsilon_{2r}-\varepsilon_{3r}-{\phi}_{r}\right)
\nonumber\\
&  -\left.  \frac{\sin^{3}\theta}{4}\sum_{r\in\Lambda}C_{r}|\alpha_{2r}%
\alpha_{4r}|\cos\left(  \varepsilon_{2r}-\varepsilon_{4r}-{\phi}_{r}\right)
\right]  ,\nonumber
\end{align}
where $\alpha_{jr}=|\alpha_{jr}|e^{i\varepsilon_{jr}}$ for $j=1,2,3,4$ and
$r\in\Lambda$. We have to maximize ${\overline{F}}_{\theta,\pi-\theta;\bot
}(\mathcal{M},\mathcal{T})$ over all possible choices of $\mathcal{M}$ and
$\mathcal{T}$, and subject to the constraints A, B, and C, that is:
\begin{equation}%
\begin{tabular}
[c]{l}%
$C_{r}>0\text{ for all }r\in\Lambda,$\\
$\sum_{j=1}^{4}|\alpha_{jr}|^{2}=1\text{ for all }r\in\Lambda,$\\
$\sum_{r\in\Lambda}C_{r}\alpha_{jr}\alpha_{kr}^{\ast}=\delta_{jk}\text{ for
all }j,k=1,2,3,4.$%
\end{tabular}
\label{bantifio}%
\end{equation}
From (\ref{bantifi}) it follows that
\begin{equation}
{\overline{F}}_{\theta,\pi-\theta;\bot}(\mathcal{M},\mathcal{T})=\frac{1}%
{2}+\frac{\sin^{3}\theta}{8}\sum_{r\in\Lambda}C_{r}\operatorname{Re}\left[
\left(  \alpha_{3r}^{\ast}-\alpha_{4r}^{\ast}\right)  \left(  \alpha
_{2r}e^{-i\phi_{r}}-\alpha_{1r}e^{i\phi_{r}}\right)  \right]  .
\label{2antifi}%
\end{equation}

Then%
\begin{align}
{\overline{F}}_{\theta,\pi-\theta;\bot}(\mathcal{M},\mathcal{T})  &  \leq
\frac{1}{2}+\frac{\sin^{3}\theta}{8}\sum_{r\in\Lambda}C_{r}|\alpha_{3r}^{\ast
}-\alpha_{4r}^{\ast}|\times|\alpha_{2r}e^{-i\phi_{r}}-\alpha_{1r}e^{i\phi_{r}%
}|\nonumber\\
&  \leq\frac{1}{2}+\frac{\sin^{3}\theta}{8}\left[  \left(  \sum_{r\in\Lambda
}C_{r}|\alpha_{3r}^{\ast}-\alpha_{4r}^{\ast}|^{2}\right)  \left(  \sum
_{r\in\Lambda}C_{r}|\alpha_{2r}^{\ast}-\alpha_{1r}^{\ast}|^{2}\right)
\right]  ^{1/2}\nonumber\\
&  =\frac{1}{2}+\frac{\sin^{3}\theta}{8}\left[  \sum_{r\in\Lambda}C_{r}%
|\alpha_{3r}|^{2}+\sum_{r\in\Lambda}C_{r}|\alpha_{4r}|^{2}-2\operatorname{Re}%
\left(  \sum_{r\in\Lambda}C_{r}\alpha_{3r}\alpha_{4r}^{\ast}\right)  \right]
^{1/2}\nonumber\\
&  \times\left[  \sum_{r\in\Lambda}C_{r}|\alpha_{1r}|^{2}+\sum_{r\in\Lambda
}C_{r}|\alpha_{2r}|^{2}-2\operatorname{Re}\left(  \sum_{r\in\Lambda}%
C_{r}\alpha_{2r}\alpha_{1r}^{\ast}e^{-2i\phi_{r}}\right)  \right]
^{1/2}\nonumber\\
&  =\frac{1}{2}+\frac{\sin^{3}\theta}{4}\left[  1-\operatorname{Re}\left(
\sum_{r\in\Lambda}C_{r}\alpha_{1r}\alpha_{2r}^{\ast}e^{2i\phi_{r}}\right)
\right]  ^{1/2}\nonumber\\
&  \leq\frac{1}{2}\left(  1+\frac{\sin^{3}\theta}{\sqrt{2}}\right)  .
\label{optanti}%
\end{align}

Let us chose
\begin{equation}%
\begin{tabular}
[c]{l}%
$\Lambda=\{1,2,3,4\},$\\
$C_{r}=1\text{ for all }r\in\Lambda,$\\
$\alpha_{1r}=\frac{e^{-2i\phi_{r}}}{2},\alpha_{2r}=-\frac{1}{2}\text{ for all
}r\in\Lambda,$\\
$\alpha_{3r}=\frac{-e^{-i\phi_{r}}}{\sqrt{2}}\text{ for }r=1,2\text{ and
}\alpha_{3r}=0\text{ for }r=3,4,$\\
$\alpha_{4r}=0\text{ for }r=1,2\text{ and }\alpha_{4r}=\frac{e^{-i\phi_{r}}%
}{\sqrt{2}}\text{ for }r=3,4,$\\
$\phi_{1}=\frac{\pi}{2},\phi_{2}=\frac{3\pi}{2},\phi_{3}=0\text{ and }\phi
_{4}=\pi.$%
\end{tabular}
\label{antichoi}%
\end{equation}
For the choice (\ref{antichoi}), one can see that all the conditions in
(\ref{bantifio}) are satisfied, then%
\begin{equation}
{\overline{F}}_{\theta,\pi-\theta;\bot}(\mathcal{M},\mathcal{T})=\frac{1}%
{2}\left(  1+\frac{\sin^{3}\theta}{\sqrt{2}}\right)  ={\overline{F}}%
_{\theta,\pi-\theta;\bot}^{\max}={\overline{F}}_{\theta,\pi-\theta;||}^{\max}.
\label{parantimax}%
\end{equation}
Expressed in terms of the basis $\{|00\rangle,|11\rangle,|01\rangle
,|10\rangle\}$, the elements of the POVM give the following matrix:
\begin{equation}
\left[
\begin{array}
[c]{cccc}%
{\alpha}_{11}=-\frac{1}{2} & {\alpha}_{21}=-\frac{1}{2} & {\alpha}_{31}%
=\frac{i\sqrt{2}}{2} & {\alpha}_{41}=0\\
{\alpha}_{12}=-\frac{1}{2} & {\alpha}_{22}=-\frac{1}{2} & {\alpha}_{32}%
=-\frac{i\sqrt{2}}{2} & {\alpha}_{42}=0\\
{\alpha}_{13}=\frac{1}{2} & {\alpha}_{23}=-\frac{1}{2} & {\alpha}_{33}=0 &
{\alpha}_{43}=\frac{1}{\sqrt{2}}\\
{\alpha}_{14}=\frac{1}{2} & {\alpha}_{24}=-\frac{1}{2} & {\alpha}_{34}=0 &
{\alpha}_{44}=-\frac{1}{\sqrt{2}}%
\end{array}
\right]  . \label{haar}%
\end{equation}
If instead of this strategy we use a POVM whose elements (expressed in terms
of the above basis) give the Fourier transorm of dimension $4$, we can find
that%
\[
{\overline{F}}_{\theta,\pi-\theta;\bot}(\mathcal{M},\mathcal{T})=\frac{1}%
{2}\left[  1+\frac{\sin^{3}\theta}{16}\left(  2\sqrt{2}\sin\left(  \frac{\pi
}{4}-\phi_{2}\right)  -4\cos\phi_{3}+2\sin\phi_{4}\right)  \right]  ,
\]
which attains the maximum value $\frac{1}{2}+\left(  \frac{\left(  3+\sqrt
{2}\right)  \sin^{3}\theta}{8}\right)  $, for $\phi_{2}=\frac{7\pi}{4}%
,\phi_{3}=\pi,\phi_{4}=\frac{\pi}{2}$ and $\phi_{1}$ arbitrary. Observe that
$\frac{1}{2}+\left(  \frac{\left(  3+\sqrt{2}\right)  \sin^{3}\theta}%
{8}\right)  <\frac{1}{2}\left(  1+\frac{\sin^{3}\theta}{\sqrt{2}}\right)  $,
the value in (\ref{parantimax}). Finally, it is important to remark that even
if
\[
\dim\mathcal{L}\left(  S_{\theta,\pi-\theta}^{||}\right)  =3<\dim
\mathcal{L}\left(  S_{\theta,\pi-\theta}^{\bot}\right)  =4,
\]
we have that ${\overline{F}}_{\theta,\pi-\theta;\bot}^{\max}={\overline{F}%
}_{\theta,\pi-\theta;||}^{\max}$. So, once again, the dimensional argument
fails. Observe that the matrix (\ref{haar}) is \textquotedblleft
similar\textquotedblright\ to
\[
\left[
\begin{array}
[c]{rrrr}%
-\frac{1}{2} & \frac{1}{2} & \frac{1}{\sqrt{2}} & 0\\
-\frac{1}{2} & \frac{1}{2} & -\frac{1}{\sqrt{2}} & 0\\
\frac{1}{2} & \frac{1}{2} & 0 & \frac{1}{\sqrt{2}}\\
\frac{1}{2} & \frac{1}{2} & 0 & -\frac{1}{\sqrt{2}}%
\end{array}
\right]  ,
\]
the Haar transform of dimension 4 (see, \emph{e.g.}, \cite{da}).

\section{LOCC protocol}

We describe here an LOCC protocol for optimally estimating the direction of a
qubit $\left\vert \psi\left(  \theta,\phi\right)  \right\rangle $ ($\theta$ is
fixed), when the supplied states are parallel states $\left\vert {\Psi}%
_{2,0}\right\rangle =|\psi(\theta,\phi)\rangle\otimes|\psi(\theta,\phi
)\rangle$. For LOCC protocols the optimality does not change whether the
supplied two qubits are parallel or anti-parallel (or anything else), as far
they are product states. The steps of the protocol are the following:

\begin{description}
\item[1] Perform the PV (projection valued) measurement $\{P[\frac{1}{\sqrt
{2}}(|0\rangle+|1\rangle)],P[\frac{1}{\sqrt{2}}(|0\rangle-|1\rangle)]\}$ on
the first qubit.

\item[2.1] If $P[\frac{1}{\sqrt{2}}(|0\rangle+|1\rangle)]$ is the outcome of
the measurement in (1) (the probability of this event being $\frac{1}%
{2}\left(  1+\sin\theta\cos\phi\right)  $), the PV measurement $\{P[\frac
{1}{\sqrt{2}}(|0\rangle+i|1\rangle)],P[\frac{1}{\sqrt{2}}(|0\rangle
-i|1\rangle)]\}$ is performed on the second qubit.

\item[2.1.1] If $P[\frac{1}{\sqrt{2}}(|0\rangle+i|1\rangle)]$ is the outcome
of the measurement in (2.1) (with probability $\frac{1}{2}\left(  1+\sin
\theta\sin\phi\right)  $), the estimated state is taken as $|\psi(\theta
,\frac{\pi}{4})\rangle$.

\item[2.1.2] If $P[\frac{1}{\sqrt{2}}(|0\rangle-i|1\rangle)]$ is the outcome
of the measurement in (2.1) (with probability $\frac{1}{2}\left(  1-\sin
\theta\sin\phi\right)  $), the estimated state is taken as $|\psi(\theta
,\frac{7\pi}{4})\rangle$.

\item[2.2] If $P[\frac{1}{\sqrt{2}}(|0\rangle-|1\rangle)]$ is the outcome of
the measurement in (1) (the probability of this event being $\frac{1}%
{2}\left(  1-\sin\theta\cos\phi\right)  $), the PV measurement $\{P[\frac
{1}{\sqrt{2}}(|0\rangle+i|1\rangle)],P[\frac{1}{\sqrt{2}}(|0\rangle
-i|1\rangle)]\}$ is performed on the second qubit.

\item[2.2.1] If $P[\frac{1}{\sqrt{2}}(|0\rangle+i|1\rangle)]$ is the outcome
of the measurement in (2.2) (with probability $\frac{1}{2}\left(  1+\sin
\theta\sin\phi\right)  $), the estimated state is taken as $|\psi(\theta
,\frac{3\pi}{4})\rangle$.

\item[2.2.2] If $P[\frac{1}{\sqrt{2}}(|0\rangle-i|1\rangle)]$ is the outcome
of the measurement in (2.2) (with probability $\frac{1}{2}\left(  1-\sin
\theta\sin\phi\right)  $), the estimated state is taken as $|\psi(\theta
,\frac{5\pi}{4})\rangle$.
\end{description}

The average fidelity is then given by
\begin{align*}
\overline{F}_{2,LOCC}(\theta)  &  =\frac{1}{2\pi}%
{\displaystyle\int\nolimits_{\phi=0}^{2\pi}}
\left\{  \left[  \frac{1}{2}\left(  1+\sin\theta\cos\phi\right)  \times
\frac{1}{2}\left(  1+\sin\theta\sin\phi\right)  \times|\langle\psi
(\theta,\frac{\pi}{4})|\psi(\theta,\phi\rangle|^{2}\right]  \right. \\
&  +\left[  \frac{1}{2}\left(  1+\sin\theta\cos\phi\right)  \times\frac{1}%
{2}\left(  1-\sin\theta\sin\phi\right)  \times|\langle\psi(\theta,\frac{7\pi
}{4})|\psi(\theta,\phi\rangle|^{2}\right] \\
&  +\left[  \frac{1}{2}\left(  1-\sin\theta\cos\phi\right)  \times\frac{1}%
{2}\left(  1+\sin\theta\sin\phi\right)  \times|\langle\psi(\theta,\frac{3\pi
}{4})|\psi(\theta,\phi\rangle|^{2}\right] \\
&  +\left.  \left[  \frac{1}{2}\left(  1-\sin\theta\cos\phi\right)
\times\frac{1}{2}\left(  1-\sin\theta\sin\phi\right)  \times|\langle
\psi(\theta,\frac{5\pi}{4})|\psi(\theta,\phi\rangle|^{2}\right]  \right\}
d\phi
\end{align*}%
\begin{align*}
&  =\frac{1}{8\pi}%
{\displaystyle\int\nolimits_{\phi=0}^{2\pi}}
\left\{  \left[  1+\sqrt{2}\sin\theta\sin\left(  \phi+\frac{\pi}{4}\right)
+\frac{\sin^{2}\theta\sin2\phi}{2}\right]  \times\left[  1-\sin^{2}\theta
\sin^{2}\left(  \frac{\phi}{2}-\frac{\pi}{8}\right)  \right]  \right. \\
&  +\left[  1-\sqrt{2}\sin\theta\sin\left(  \phi-\frac{\pi}{4}\right)
-\frac{\sin^{2}\theta\sin2\phi}{2}\right]  \times\left[  1-\sin^{2}\theta
\sin^{2}\left(  \frac{\phi}{2}-\frac{7\pi}{8}\right)  \right] \\
&  +\left[  1+\sqrt{2}\sin\theta\sin\left(  \phi-\frac{\pi}{4}\right)
-\frac{\sin^{2}\theta\sin2\phi}{2}\right]  \times\left[  1-\sin^{2}\theta
\sin^{2}\left(  \frac{\phi}{2}-\frac{3\pi}{8}\right)  \right] \\
&  +\left.  \left[  1-\sqrt{2}\sin\theta\sin\left(  \phi+\frac{\pi}{4}\right)
+\frac{\sin^{2}\theta\sin2\phi}{2}\right]  \times\left[  1-\sin^{2}\theta
\sin^{2}\left(  \frac{\phi}{2}-\frac{5\pi}{8}\right)  \right]  \right\}
d\phi\\
&  =\frac{1+\cos^{2}\theta}{2}+\frac{\sin^{3}\theta}{2\sqrt{2}}.
\end{align*}
The value obtained is then equal to$\overline{F}_{2,0}^{\max}(\theta)$ (where
all types of measurements are allowed). Thus%
\[%
\begin{tabular}
[c]{lll}%
$\overline{F}_{2,LOCC}^{\max}(\theta)=\overline{F}_{2,0}^{\max}(\theta
)=\frac{1+\cos^{2}\theta}{2}+\frac{\sin^{3}\theta}{2\sqrt{2}},$ &  & for every
$\theta\in\lbrack0,\pi].$%
\end{tabular}
\]
The optimal fidelity obtained by performing LOCC is then equal to the optimal
fidelity obtained by performing joint measurements on parallel qubits.

In the attempt to generalize the above analysis to $N$ parallel qubits, there
is some evidence that von Neumann measurements on individual qubits may not
achieve the optimal fidelity. A POVM (on the $N$-th qubit) consisting of
$2^{N-2}$ rank-one elements may in fact give better fidelity \cite{w}.

\section{Discussion}

Gisin and Popescu \cite{gisinpopescu99} have shown that more information about
the direction of (the Bloch vector of) a qubit $|\psi(\theta,\phi)\rangle
=\cos(\theta/2)|0\rangle+e^{i\phi}\sin(\theta/2)|1\rangle$ can be obtained
from anti-parallel states $|\Psi_{1,1}(\theta,\phi)\rangle=|\psi(\theta
,\phi)\rangle\otimes|\psi(\pi+\theta,\pi-\phi)\rangle$, compared to parallel
states $|\Psi_{2,0}(\theta,\phi)\rangle=|\psi(\theta,\phi)\rangle\otimes
|\psi(\theta,\phi)\rangle$, where $(\theta,\phi)$ is uniformly distributed
over $[0,\pi]\times\lbrack0,2\pi)$. They attributed the cause of this fact to
the difference between the dimensions of the subspaces spanned by parallel and
anti-parallel states respectively.

When $\theta=\pi/2$, there is no difference in the amount of information as in
that case (and only in that case) exact spin-flipping is possible. For any
fixed $\theta$, the dimension of the space spanned by $N$ and $M$ qubits
respectively identical and orthogonal to $|\psi(\theta,\phi)\rangle$ is
$(N+M+1)$. We found that, whenever we fix $\theta\neq0$, $\neq\pi/2$ or
$\neq\pi$, anti-parallel states always give more information about the
direction of the qubit. We generalized this to the case of $N$ and $M$ qubits
respectively identical and orthogonal to $|\psi(\theta,\phi)\rangle$. Here the
measurement basis for the optimal estimation strategy always turns out to be
the Fourier basis.

We considered the case of two diametrically opposite circles. We found that
both the sets $\{|\Psi_{1,1}(\theta,\phi)\rangle:\phi\in\lbrack0,2\pi
)\}\cup\{|\Psi_{1,1}(\pi-\theta,\pi+\phi)\rangle:\phi\in\lbrack0,2\pi)\}$ and
$\{|\Psi_{2,0}(\theta,\phi)\rangle:\phi\in\lbrack0,2\pi)\}\cup\{|\Psi
_{2,0}(\pi-\theta,\pi+\phi)\rangle:\phi\in\lbrack0,2\pi)\}$ give the same
information about the direction of $|\psi(\theta,\phi)\rangle$, even though
the linear span of the first set is four and the one of the second set is
three. The scenario described is not exactly phase estimation, still the
Fourier basis is again the optimal measurement basis for the case of parallel
qubits. This does not hold for anti-parallel qubts.

We have seen that encoding of the two qubits to parallel or anti-parallel
states is nothing special in regard to optimal extraction of information about
the direction of the qubit $|\psi(\theta,\phi)\rangle$ from a fixed circle. In
particular we have seen that the econding of $|\psi(\theta,\phi)\rangle$ in
the form $|\psi(\theta,\phi)\rangle\otimes|\psi(\theta+\theta_{0},\phi
)\rangle$, where $\theta_{0}$ is fixed in $[-\theta,\pi-\theta]$, can provide
more information compared to the case of anti-parallel qubits, when
$\theta_{0}\neq0$, $\neq\pi/2$ or $\neq\pi$.

When two parallel qubits are supplied from a circle, we have verified that a
measurement strategy using LOCC can give rise to optimal information. This is
interesting from the experimental point of view since it is practically
difficult to perform measurements in an entangled basis (see, \emph{e.g.},
\cite{z}).

Massar \cite{massar00} has shown that in the case of extracting information
about the direction of a qubit taken from a uniform distribution over whole
the Bloch sphere, even parallel qubits can give better information compared to
anti-parallel qubits provided one choses the proper score. From the point of
view of estimation of statistical parameters, this argument is of course
reasonable. This does not shed light on the reason whether there is some
physical connection between impossibility of spin-flipping and outperformance
of anti-parallel over parallel qubits. Moreover it is not clear what kind of
score is preferable, given some supplied multiqubit states; even though the
physically motivated score should be the one which directly estimates the
direction of the qubit. (We considered this score.)

\bigskip

We conclude with some open problems:

\begin{itemize}
\item Determine which one of the following two sets provides more information
about the direction of $|\psi(\theta,\phi)\rangle$: $\{|\Psi_{n,N-n}%
(\theta,\phi)\rangle:\phi\in\lbrack0,2\pi)\}\cup\{|\Psi_{n,N-n}(\pi-\theta
,\pi+\phi)\rangle:\phi\in\lbrack0,2\pi)\}$ and $\{|\Psi_{N,0}(\theta
,\phi)\rangle:\phi\in\lbrack0,2\pi)\}\cup\{|\Psi_{N,0}(\pi-\theta,\pi
+\phi)\rangle:\phi\in\lbrack0,2\pi)\}$.

\item Given any $\theta\in\lbrack0,\pi]$, determine for which values of
$\theta_{0}\in\lbrack-\theta,\pi-\theta]$, which one of the following sets
provides more information about the direction of $|\psi(\theta,\phi)\rangle$:
$\{|\psi(\theta,\phi)\rangle^{\otimes n}\otimes|\psi(\theta+\theta_{0}%
,\phi)\rangle^{\otimes(N-n)}:\phi\in\lbrack0,2\pi)\}$.

\item Given $\theta\in\lbrack0,\pi]$, determine whether a strategy using LOCC
is optimal for estimating the direction of $|\psi(\theta,\phi)\rangle$, when
the supplied state belongs to the set $\{|\Psi_{N,0}(\theta,\phi)\rangle
:\phi\in\lbrack0,2\pi)\}$. If this is true, what is then the corresponding LOCC?
\end{itemize}

\bigskip

\noindent\textbf{Acknowledgement.} SLB currently holds a Royal Society-Wolfson
Research Merit award. This work is funded in part by EPSRC grants GR/87406 and
GR/S56252. SG would like to thank Emili Bagan and Ramon Mu\~{n}oz-Tapia for
useful discussion regarding this paper and for indicating the reference
\cite{bagan02}. Part of this work has been done while the authors where at The
Isaac Newton Institute for Mathematical Sciences (Cambridge, United Kingdom),
during the Quantum Information Science Programme (16 Aug - 17 Dec 2004).

\end{document}